\documentclass[]{aastex701} 
\usepackage{graphicx}	
\usepackage{amsmath}	

\usepackage{longtable}
\usepackage{subfig}
\usepackage{booktabs}
\usepackage{rotating}
\usepackage{CJK}
\usepackage{xcolor}

\begin{document}

\title[Eclipsed Bursts from SGR J1935+2154]{Eclipsed X-ray Bursts from Magnetar SGR J1935+2154 and the Fireball Measurements}
\shortauthors{Xie et al.}

\correspondingauthor{Aming Chen}
\email{chensam@ccnu.edu.cn}
\correspondingauthor{Yun-Wei Yu}
\email{yuyw@ccnu.edu.cn}
\correspondingauthor{Shao-Lin Xiong}
\email{xiongsl@ihep.ac.cn}

\author[0000-0001-9217-7070]{Sheng-Lun Xie}
\affiliation{Institute of Astrophysics, Central China Normal University, Wuhan 430079, China}
\affiliation{State Key Laboratory of Particle Astrophysics, Institute of High Energy Physics, Chinese Academy of Sciences, 19B Yuquan Road, Beijing 100049, China}
\email{xiesl@mails.ccnu.edu.cn}

\author[0000-0003-1838-8456]{Aming Chen}
\affiliation{Institute of Astrophysics, Central China Normal University, Wuhan 430079, China}
\affiliation{Education Research and Application Center, National Astronomical Data Center, Wuhan 430079, China}
\email{chensam@ccnu.edu.cn}

\author[0000-0002-1067-1911]{Yun-Wei Yu}
\affiliation{Institute of Astrophysics, Central China Normal University, Wuhan 430079, China}
\affiliation{Education Research and Application Center, National Astronomical Data Center, Wuhan 430079, China}
\email{yuyw@ccnu.edu.cn}

\author[0000-0002-4771-7653]{Shao-Lin Xiong}
\affiliation{State Key Laboratory of Particle Astrophysics, Institute of High Energy Physics, Chinese Academy of Sciences, 19B Yuquan Road, Beijing 100049, China}
\email{xiongsl@ihep.ac.cn}

\author[0000-0001-7584-6236]{Hua Feng}
\affiliation{State Key Laboratory of Particle Astrophysics, Institute of High Energy Physics, Chinese Academy of Sciences, 19B Yuquan Road, Beijing 100049, China}
\email{hfeng@ihep.ac.cn}

\author[0000-0001-5586-1017]{Shuang-Nan Zhang}
\affiliation{State Key Laboratory of Particle Astrophysics, Institute of High Energy Physics, Chinese Academy of Sciences, 19B Yuquan Road, Beijing 100049, China}
\affiliation{University of Chinese Academy of Sciences, Beijing 100049, China}
\email{zhangsn@ihep.ac.cn}

\author[0000-0002-7835-8585]{Zi-Gao Dai}
\affiliation{Department of Astronomy, School of Physical Sciences, University of Science and Technology of China, Hefei 230026, China}
\email{daizg@ustc.edu.cn}

\author[0000-0001-8664-5085]{Wang-Chen Xue}
\affiliation{State Key Laboratory of Particle Astrophysics, Institute of High Energy Physics, Chinese Academy of Sciences, 19B Yuquan Road, Beijing 100049, China}
\affiliation{University of Chinese Academy of Sciences, Beijing 100049, China}
\email{xuewangchen@ihep.ac.cn}

\author{Ming-Yu Ge}
\affiliation{State Key Laboratory of Particle Astrophysics, Institute of High Energy Physics, Chinese Academy of Sciences, 19B Yuquan Road, Beijing 100049, China}
\email{gemy@ihep.ac.cn}

\author[0000-0003-4585-589X]{Xiao-Bo Li}
\affiliation{State Key Laboratory of Particle Astrophysics, Institute of High Energy Physics, Chinese Academy of Sciences, 19B Yuquan Road, Beijing 100049, China}
\email{lixb@ihep.ac.cn}

\author[0000-0002-8708-0597]{Liang-Duan Liu}
\affiliation{Institute of Astrophysics, Central China Normal University, Wuhan 430079, China}
\affiliation{Education Research and Application Center, National Astronomical Data Center, Wuhan 430079, China}
\email{liuld@mail.ccnu.edu.cn}

\author[0009-0004-1887-4686]{Jia-Cong Liu}
\affiliation{State Key Laboratory of Particle Astrophysics, Institute of High Energy Physics, Chinese Academy of Sciences, 19B Yuquan Road, Beijing 100049, China}
\affiliation{University of Chinese Academy of Sciences, Beijing 100049, China}
\email{liujc98@ihep.ac.cn}

\author[0009-0006-5506-5970]{Wen-Jun Tan}
\affiliation{State Key Laboratory of Particle Astrophysics, Institute of High Energy Physics, Chinese Academy of Sciences, 19B Yuquan Road, Beijing 100049, China}
\affiliation{University of Chinese Academy of Sciences, Beijing 100049, China}
\email{tanwj@ihep.ac.cn}

\author[0009-0008-8053-2985]{Chen-Wei Wang}
\affiliation{State Key Laboratory of Particle Astrophysics, Institute of High Energy Physics, Chinese Academy of Sciences, 19B Yuquan Road, Beijing 100049, China}
\affiliation{University of Chinese Academy of Sciences, Beijing 100049, China}
\email{cwwang@ihep.ac.cn}

\author{Shu-Xu Yi}
\affiliation{State Key Laboratory of Particle Astrophysics, Institute of High Energy Physics, Chinese Academy of Sciences, 19B Yuquan Road, Beijing 100049, China}
\email{sxyi@ihep.ac.cn}

\author{Peng Zhang}
\affiliation{State Key Laboratory of Particle Astrophysics, Institute of High Energy Physics, Chinese Academy of Sciences, 19B Yuquan Road, Beijing 100049, China}
\affiliation{College of Electronic and Information Engineering, Tongji University, Shanghai 201804, China}
\email{zhangp97@ihep.ac.cn}

\author[0000-0001-5348-7033]{Yan-Qiu Zhang}
\affiliation{State Key Laboratory of Particle Astrophysics, Institute of High Energy Physics, Chinese Academy of Sciences, 19B Yuquan Road, Beijing 100049, China}
\affiliation{University of Chinese Academy of Sciences, Beijing 100049, China}
\email{zhangyanqiu@ihep.ac.cn}

\author{Zhen Zhang}
\affiliation{State Key Laboratory of Particle Astrophysics, Institute of High Energy Physics, Chinese Academy of Sciences, 19B Yuquan Road, Beijing 100049, China}
\email{zhangzhen@ihep.ac.cn}

\author[0009-0001-7226-2355]{Chao Zheng}
\affiliation{State Key Laboratory of Particle Astrophysics, Institute of High Energy Physics, Chinese Academy of Sciences, 19B Yuquan Road, Beijing 100049, China}
\affiliation{University of Chinese Academy of Sciences, Beijing 100049, China}
\email{zhengchao97@ihep.ac.cn}

\author[0000-0001-8868-4619]{Xiao-Ping Zheng}
\affiliation{Institute of Astrophysics, Central China Normal University, Wuhan 430079, China}
\affiliation{Education Research and Application Center, National Astronomical Data Center, Wuhan 430079, China}
\email{zhxp@ccnu.edu.cn}



\begin{abstract}
In theory, burst activity of the magnetar can lead to the formation of fireballs trapped by the magnetic field and corotating with the star. However, the smoking-gun observational evidence of the fireball is elusive. 
We envisage that the fireball emission should occasionally be eclipsed by the magnetar, especially when the burst duration is comparable to the magnetar's spin period. 
In this work, we first discover a peculiar type of burst whose light curve has a plateau-like feature among the long bursts of the magnetar SGR J1935+2154 detected by GECAM and Fermi/GBM. Then, based on these bursts, we identified four burst candidates with eclipse-like characteristics. By fitting their light curves with the eclipse fireball model, the viewing angle of the magnetar relative to its spin axis is estimated to be $17^\circ \pm 10^\circ$, and the distances from the fireballs to the magnetar are found to be more than 5 times the magnetar's radius, indicating that the fireballs are suspended in the magnetosphere rather than adhering to the magnetar surface. Furthermore, we find that this configuration is well consistent with the implication of the cyclotron resonance scattering feature we found in their spectra.
Our results suggest that some intermediate X-ray bursts may originate from magnetic reconnection within the magnetosphere rather than the starquake.
\end{abstract}


\section{Introduction}\label{sec:intro}
Magnetars are a subclass of neutron stars with extremely high magnetic fields ($10^{14}\sim10^{15}$ G), which usually exhibit as pulsars/bursters of long spin periods ($2\sim10$ s) and high spin-down rates ($10^{-13}\sim10^{-11}\ {\rm s}\cdot {\rm s}^{-1}$) in our Galaxy \citep{Woods2006csxs,Kaspi2010PNAS,Mereghetti2013BrJPh,Olausen2014,KaspiBeloborodov2017ARAA,NegroYounes2024FrASS}. These objects are well-known for their bursting and flaring activity, during which the luminosity of the neutron stars rises suddenly in X-rays and soft gamma-rays.
The bright X-ray/gamma-ray transients from magnetars are usually classified into different subtypes in terms of their timescales and energetics, including ordinary short bursts, intermediate burst/flares, and giant flares \citep{Olive2004ApJ,WoodsThompson2006csxs,Mereghetti2008AARv,Israel2008ApJ}.
The short bursts are most common and typically have a duration of $0.1$--$1\,\mathrm{s}$ and a peak luminosity in the range of $10^{39}$--$10^{41}\,\mathrm{erg\,s^{-1}}$.
The giant flares are rare events, generally start with a short ($0.1$--$0.2\,\mathrm{s}$) spike of hard X-rays with peak luminosity ($10^{44}$--$10^{47}\,\mathrm{erg\,s^{-1}}$) followed by a softer pulsating tail lasting a few minutes. So far, only a total number of three giant flares have been discovered in our Galaxy, while some new candidates have recently been claimed to be detected from extragalactic sources (e.g., GRB 231115A from M82, GRB 070222 from M83, and GRB 180128A and GRB 200415A from NGC 253, \citealt{Mereghetti2024Natur,Trigg2024AA,Roberts2021Natur,Burns2021ApJ}). The intermediate bursts/flares exhibit properties with durations, luminosities, and releasing energies between short bursts and giant flares. 
Moreover, they typically have abrupt onset and endpoints, with duration timescales shorter than the spin periods.
Besides, in some so-called outburst events, the X-ray emission of magnetars can suddenly rise to be $10$--$10^3$ times brighter than their quiescent level and subsequently decay in weeks to months/years as a power-law or exponential behavior (e.g., \citealt{CotiZelati2018MNRAS}). The total energy released from an outburst is typically in the range of $10^{41}$--$10^{43}\,\mathrm{erg}$. The onsets of outbursts are usually associated with one or more ordinary bursts, but their precise relationship is still unclear.

The prevailing explanation for magnetar bursts/flares is that large amounts of energy stored in magnetic fields are released owing to a rapid change in the magnetic field structure and strength, since the rotational energy of magnetars is always too low to account for the observations. Such a scenario is strongly supported by the positive correlation between the X-ray luminosity of magnetars and the strength of their magnetic fields \citep{Thompson2000ApJ,Thompson2002ApJ,Uzdensky2011SSRv,Parfrey2013ApJ}. Magnetic fields distributed in the magnetosphere of magnetars are believed to be more complicated (e.g., twisted) than those of normal pulsars \citep{Thompson1995MNRAS,Thompson1998PhRvD,Thompson2002ApJ}, even though the global structure of the fields is still generally dipolar. Therefore, the magnetar surface can be persistently heated owing to the gradual untwisting of the field lines, resulting in a higher temperature than normal pulsars. The most concerning question is what physical processes trigger the burst activities. Some plausible mechanisms have been proposed, such as magnetic-stress-induced starquakes \citep{Thompson1995MNRAS,Jones2003ApJ,Levin2012MNRAS} and magnetic field reconnections of some ropes in the magnetosphere \citep{Duncan1992ApJ,Masada2010PASJ,Yu2011ApJ,Yu2013ApJ,Meng2014ApJ}, which lead to sudden changes in the configuration of either internal or external magnetic fields.

\begin{figure}[htbp]
\centering
\centerline{\includegraphics[width=0.45\textwidth]{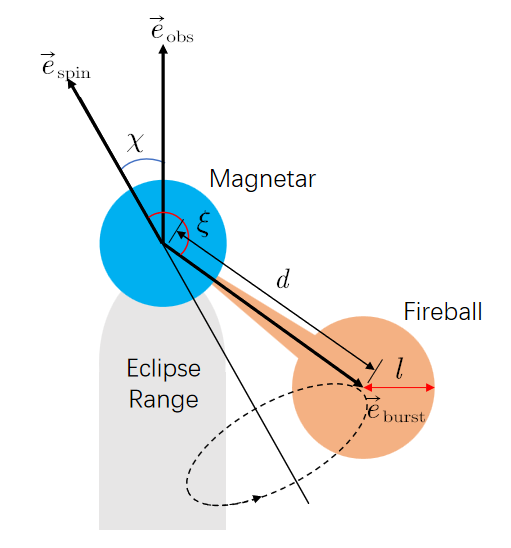}}
\caption{Illustration of the eclipse geometry of a fireball corotating with the magnetar (not to scale). Note that the corotating spherical fireball is an effective emitting region for the eclipse geometry, not the physical shape of the whole fireball plasma.}\label{fig:geometry}
\end{figure}

Observational constraints on the radiation mechanisms of magnetar bursts are usually derived from their spectral information \citep[e.g.,][]{Gogus1999ApJ,Gogus2000ApJ,vanderHorst2012ApJ}. In contrast, the light curves of bursts are usually ignored because of their complexity, except for the afterglow emission of some giant flares and outbursts \citep{Woods2001ApJ,Lyubarsky2002ApJ,Gaensler2005Nature,Cameron2005Nature,LiBeloborodov2015ApJ}. The emission region of a giant flare (probably a fireball consisting of hot electron/positron plasma) is usually suggested to be near the stellar surface and even capable of engulfing the entire magnetar \citep{Thompson1995MNRAS}. Therefore, the temporal evolution of the tail emission of flares can show a periodic modulation at the spin period as the fireball corotates with the magnetar. The modulations are believed to arise from the inhomogeneous distribution of the fireball relative to the magnetar. However, for relatively normal bursts, including the short and intermediate ones, the corresponding fireballs could be much smaller (if they are still near the stellar surface) and, in any case, only located on one side of the magnetar. Therefore, we suggest here that, for these fireballs, we would have a chance to discover the eclipse signal from their light curves if their duration can be comparable to the magnetar's spin period. As illustrated in Figure \ref{fig:geometry}, the fireball emission could be obscured by the magnetar for a period when the fireball moves into the shadow of the magnetar. Such a situation, if discovered, particularly from intermediate bursts, would enable us to constrain the fireball's size and its geometrical relationship with the magnetar and would further help distinguish the different burst-triggering mechanisms.

In this work, we aim to search for eclipse-like features in the X-ray bursts of the prolific magnetar SGR J1935+2154 and develop a geometric model to interpret these light curves. Using GECAM and Fermi/Gamma-ray Burst Monitor (GBM) data, we identify several long bursts exhibiting eclipse-like features and perform eclipse model fits to constrain the viewing angle, fireball size, and distance to the magnetar. These results provide new insights into the spatial configuration of magnetar fireballs and their possible origin. 

The structure of this paper is as follows: Section \ref{sec:eclipse_sign} describes the eclipse X-ray bursts from SGR J1935+2154, including the data analysis and model fittings. Sections \ref{sec:dicuss} and \ref{sec:summary} present the discussion and summary of our results, respectively. The details of data reduction and spectral analysis are presented in Appendices \ref{appxsec:data_reduct} and \ref{appxsec:spec_analys}, respectively.

\section{Eclipse X--ray bursts}\label{sec:eclipse_sign}

This work is focused on the observations of the magnetar SGR J1935+2154, which is one of the most active magnetars \citep[e.g.,][]{Israel2016MNRAS,Lin2020apj,Lin2020apjl,Younes2020ApJ,Borghese2022MNRAS,Cai2022a} and, in particular, has been found to host mysterious fast radio bursts \citep[FRBs; e.g.,][]{Bochenek2020nat,CHIMEFRB2020nat,Li2021NatAs,Ridnaia2021NatAs}. Since the first discovery by Swift/Burst Alert Telescope (Swift/BAT) in 2014 \citep{Stamatikos2014gcn}, SGR J1935+2154 has attracted a great number of observations, including the monitors of Fermi Gamma-ray Burst Monitor (Fermi/GBM) and Gravitational wave high-energy Electromagnetic Counterpart All-sky Monitor \citep[GECAM;][]{Lin2020apj,Lin2020apjl,Kaneko2021ApJ,Xie2022mnras,Rehan2023apj,Rehan2024ApJ,Xie2025ApJS,Rehan2025ApJS}. Up to the end of 2022, GECAM and Fermi/GBM have detected 256 and 440 bursts, respectively, resulting in a total of 627 bursts.\footnote{Note that there are duplicated bursts that are simultaneously detected by GECAM and Fermi/GBM.}
Most of these bursts are typical short bursts with a single or multiple peaks. Some statistical studies have been conducted on the distributions of bursts' energy, duration, and waiting times, from which a particularity of the burst activity hosting the FRB emission has been found \citep{Xie2024ApJ,Rehan2025ApJS}.

\begin{figure}[htbp]
\centering
\centerline{\includegraphics[width=0.5\textwidth]{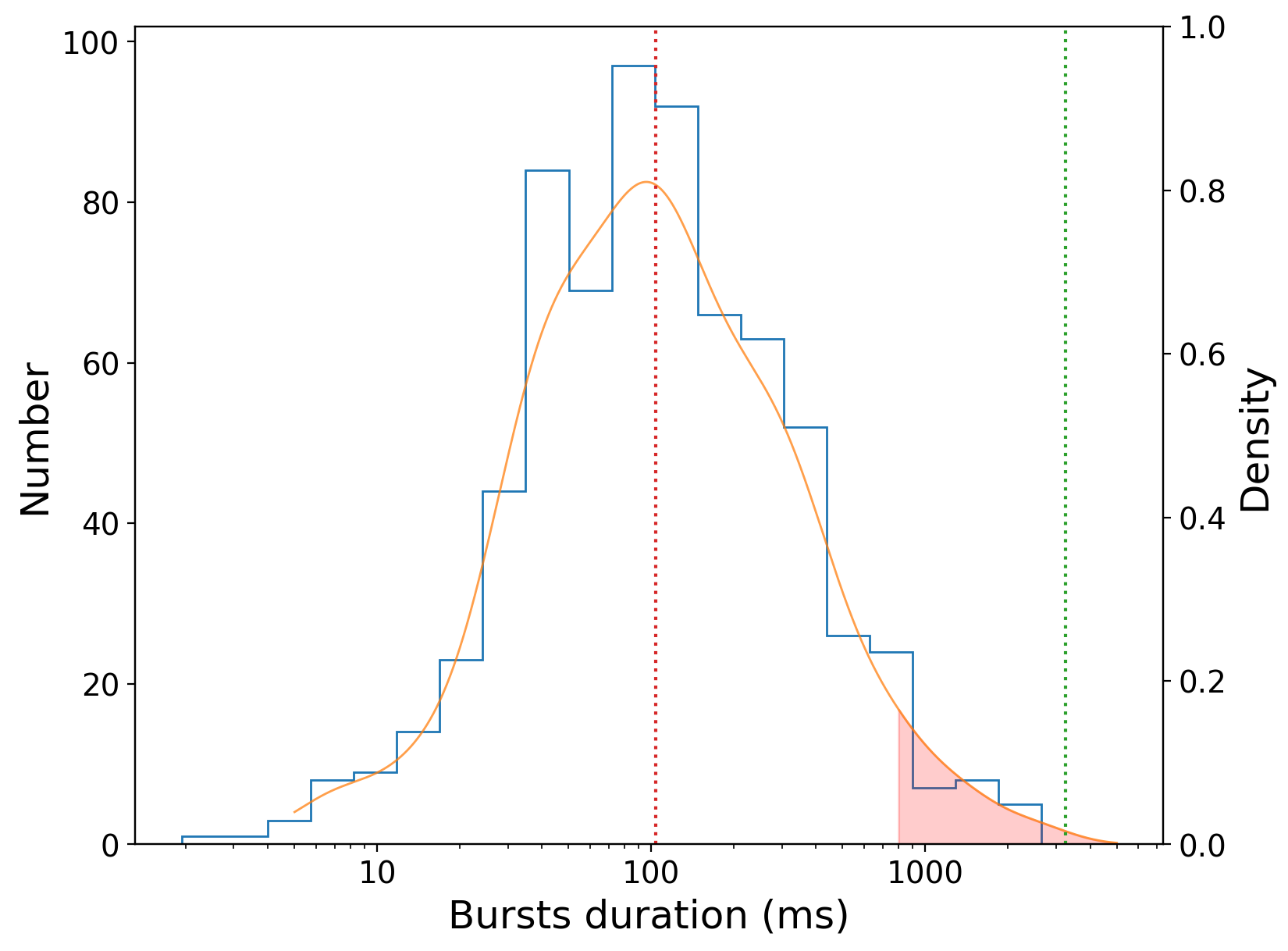}}
\caption{The histogram (blue line) and kernel density (orange line) of the duration distribution of the 627 X-ray bursts from SGR J1935+2154 detected by GECAM \citep{Xie2022mnras,Xie2025ApJS} and Fermi/GBM \citep{Lin2020apj,Lin2020apjl,Kaneko2021ApJ,Rehan2023apj,Rehan2024ApJ} up to the end of 2022. The red and green vertical dotted lines represent the mean value ($103.51$ ms) of burst duration and the spin period ($3.24$ s) of SGR J1935+2154, respectively. The duration range in which we search for eclipsed X-ray burst candidates is highlighted in the shaded region. }\label{fig:duration_dis}
\end{figure}

In Figure \ref{fig:duration_dis}, we display the duration distribution of the total 627 bursts detected by GECAM and Fermi/GBM, which determines a mean value of approximately 100 ms.
Considering the spin period of SGR J1935+2154 with $P\sim 3.24\,\mathrm{s}$, we tentatively take 800 ms (which is approximately one-fourth of the spin period) as a criterion and obtained 23 bursts above this line. Among these long bursts, eclipsing light-curve candidates are further selected based on a visual inspection of their light-curve morphologies.
Unfortunately, the light curves of these relatively long bursts often exhibit multiple peaks, making it difficult to determine whether they are modulated by the eclipse effect. 
Here we propose a method to explore the eclipse effect as follows.
Firstly, find out whether there is a specific light-curve pattern that could be intrinsic to a group of bursts. Secondly, search for eclipse signals in bursts that differ only partially from the supposed standard pattern.

\subsection{Data Analysis}
For GECAM and GBM, detectors with source angles $\leqslant50^\circ$ were selected, and the energy channels affected by the absorption edges ($30$--$40\,\mathrm{keV}$) were excluded. Further details of the data reduction procedures are provided in Appendix \ref{appxsec:data_reduct}.
Based on the cleaned and calibrated data, we then carry out time-resolved spectral analysis to explore the spectral evolution of the selected long bursts.
Background spectra are estimated from pre- and post-burst intervals using low-order polynomial fits.
The spectra are grouped to ensure sufficient counts per channel and fitted using Bayesian estimation. 
Further details of the spectral analysis, including the fitting methods and model comparisons, are summarized in Appendix \ref{appxsec:spec_analys}.

\begin{figure*}[htbp]
\centering
{\label{subfig:plateau_A}\includegraphics[width=0.45\textwidth]{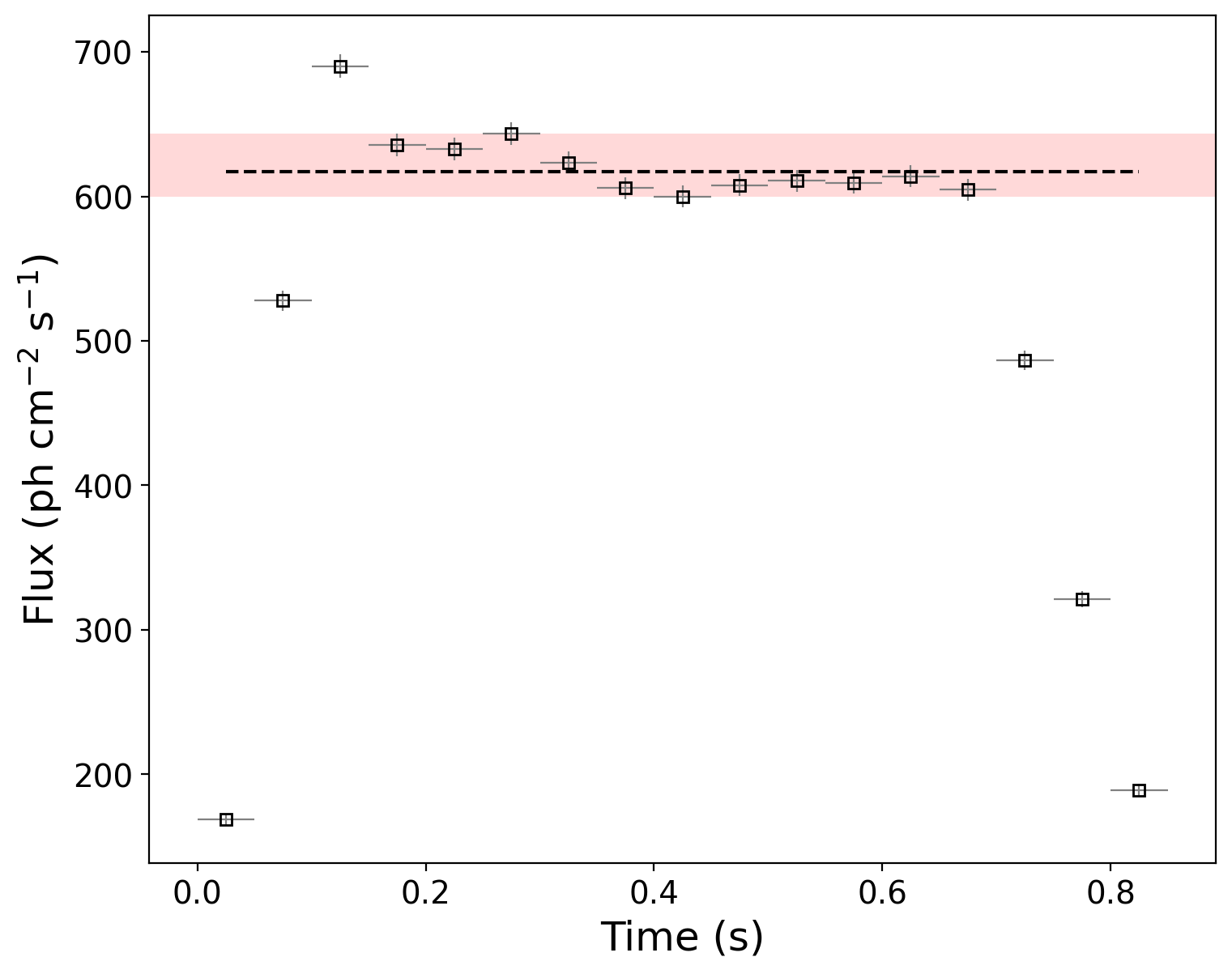}}
\hfill
{\label{subfig:ts_A}\includegraphics[width=0.45\textwidth]{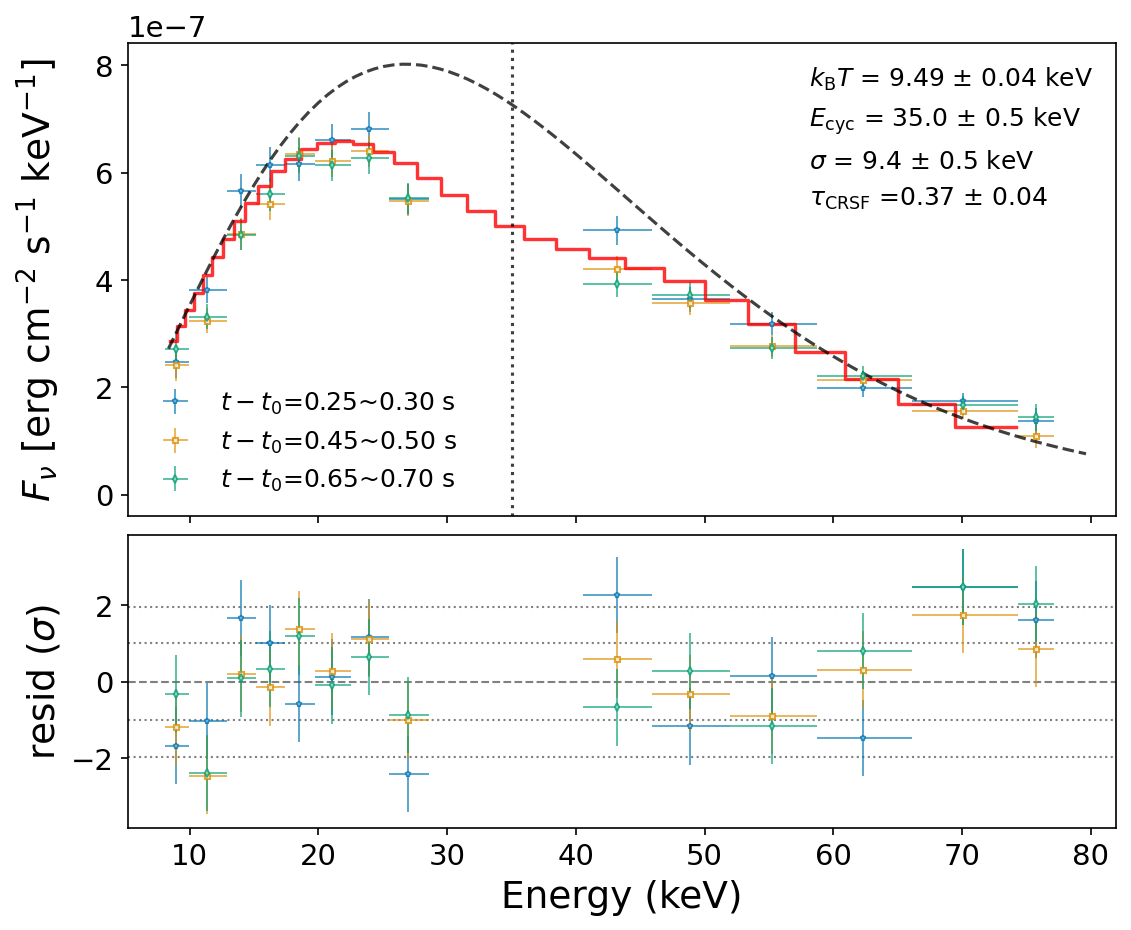}}
\hfill
\caption{
{\it Left}: the light curve of a plateau burst observed by Fermi/GBM at 2016-06-26 13:54:30.720 (labeled Burst A), where the dashed line represents the average of the plateau flux, and the fluctuation, as shown by the shaded band, is no more than $\sim6$\% of the average.
{\it Right}: the time-resolved spectra of Burst A, which can be fitted by a nonevolving BB with a CRSF absorption (red solid histogram). The residuals of the spectral fit are plotted in the lower panel. The dashed line represents the unabsorbed BB. The center of the CRSF line is labeled by the vertical dotted line, while the optical depth $\tau_{\rm CRSF}$ and width $\sigma$ of the absorption are given in the legend.
}\label{fig:lc_each_plateau}
\end{figure*}

Based on visual inspection of the intermediate bursts of SGR J1935+2154, we have indeed found several bursts whose light curves exhibit a plateau-like feature. The most representative one is the burst occurring at $t_0$ = 2016-06-26T13:54:30.720 (labeled as Burst A), as shown in the left panel of Figure \ref{fig:lc_each_plateau}. 
It exhibits the clearest and stablest plateau profile among the identified long bursts.
Such a plateau indicates that the emission region of this burst likely has a nearly fixed configuration, i.e., the fireball is in dynamic equilibrium. 

To obtain the emitting areas of fireballs, we tested several spectral models, including a single-blackbody (BB) model, a two-BB (BB + BB) model, and a BB plus cyclotron resonant scattering feature (BB + CRSF) model. For the representative plateau burst (Burst A), the temperature of a single-BB model fluctuates slightly around a constant value, indicating that the slight differences among the spectra could stem from the dynamic fluctuation of the steady fireball. However, a clear residual dip appears at $20$--$50\,\mathrm{keV}$ (see the Figure \ref{appxfig:4spec_model_A} in Appendix \ref{appxsec:spec_analys}), indicating the presence of an absorption feature.
Adopting a unified temperature at $k_\mathrm{B}T=9.49$ keV and introducing a CRSF component at $E_\mathrm{cyc}=35.0$ keV significantly improves the fit (see the Table \ref{appxtab:5burst_spec_resul} in Appendix \ref{appxsec:spec_analys}), suggesting a magnetic field strength much lower than the surface value of the magnetar, i.e., $\sim3\times10^{12}$ G.
This indicates that the line-forming region is far away from the surface. Generally speaking, in addition to the BB component, wider-band spectra of X-ray bursts of magnetars sometimes have another spectral component arising from the Comptonization of the thermal photons \citep{Ioka2020ApJ,Yang2021ApJ,Yamasaki2022MNRAS,Zhang2023MNRAS}, usually dominating the higher energy band ($\gtrsim 80$ keV). This component is not considered here.
In summary, from a qualitative perspective, the BB + CRSF model provides a better statistical and physical description of the data, as supported by the Akaike Information Criterion (AIC) and Bayesian Information Criterion (BIC) tests.
Therefore, the plateau emission, with a nearly constant temperature and emission area, undoubtedly provides a unique opportunity to explore the eclipse effect. 

Taking Burst A as a template, we visually inspected the remaining long bursts and identified four bursts showing similar behaviors to that of Burst A, but with a partial deletion, labeled as Bursts B--E.
The light curves of these four bursts are shown in Figure \ref{fig:lc_each_eclipsed}, where the rapid rise and decline phases are not fully displayed to more clearly show the details of the plausible "plateau phase". The spectral fitting of these bursts reveals a non-evolving temperature but a varying emission area, which can be naturally explained by the eclipse effect owing to the emission fireball moving into the shadow of the magnetar. Therefore, we take these four bursts as samples of eclipsed X-ray bursts in this work.

\subsection{Eclipse Modeling}\label{sec:eclipse_model}

Based on the scenario illustrated in Figure \ref{fig:geometry}, we developed a magnetar-fireball eclipse model to fit the light curves of selected bursts. As a zeroth-order approximation, we simply treat the emitting regions of bursts as a spherical fireball corotating with the magnetar, although their realistic geometrical structure can be much more complicated. 
We introduce a reference frame with the $z$-axis toward the observer, i.e., $\vec{e}_{\mathrm{obs}}=(0,0,1)$, and the $x$-axis in the plane of the observer and the magnetar's spin axis. Therefore, the unit vectors of the spin axis and the fireball can be given by
\begin{eqnarray}
\vec{e}_{\mathrm{spin}}&=&\left( \sin \chi ,0,\cos \chi \right), \\
\vec{e}_{\mathrm{burst}}&=&\left( \sin \chi \cos \xi +\cos \chi \sin \xi \cos \gamma ,\sin \xi \sin \gamma ,\cos \chi \cos \xi -\sin \chi \sin \xi \cos \gamma \right),
\end{eqnarray}
where $\chi$ is the angle between the observer and the spin axis and $\xi$ is the angle between the spin axis and the fireball. Since the fireball corotates with the magnetar, the vector $\vec{e}_{\mathrm{burst}}$ rotates around $\vec{e}_{\mathrm{spin}}$. Therefore, the rotational phase is given by $\gamma \left( t \right) =\Omega t+\phi $, with $\Omega=2\pi/P$ and $\phi$ being the azimuth angle of the fireball, which is related to the choice of starting time during each burst.

We denote $\theta$ as the angle between the fireball and the observer (i.e., $\cos\theta=\vec{e}_{\mathrm{burst}}\cdot
\vec{e}_{\mathrm{obs}}$), which varies from $|\chi-\xi|$ to $|\chi+\xi|$ as the fireball corotates with the magnetar. For a given pair of angles $\chi$ and $\xi$ that satisfy with $|\chi+\xi|\leq\pi/2$ or $|\chi+\xi|\geq 3\pi/2$, the magnetar itself would not eclipse the fireball emission, and the emitting area is given by $A_{\mathrm{obs}}=\pi l^2$, where $l$ is the radius of the fireball.
The eclipse process occurs when $\pi/2<|\chi+\xi|<3\pi/2$, and the visible area is determined by the apparent distance between the fireball and the magnetar $\rho =d\cos \left( \theta -\pi /2 \right) $ as
\begin{equation}
A_{\text{obs}}=\left\{
\begin{array}{ll}
\pi l^2,&\rho>l+R_{\mathrm{ns}},\\
\pi l^2-\Delta A_1-\Delta A_2,&(l+R_{\mathrm{ns}})>\rho>\sqrt{l^2-R_{\mathrm{ns}}^2},\\
\pi l^2-\pi R_{\mathrm{ns}}^2+\Delta A_2-\Delta A_1,&\sqrt{l^2-R_{\mathrm{ns}}^2}>\rho>(l-R_{\mathrm{ns}}),\\
\pi l^2-\pi R_{\mathrm{ns}}^2,&(l-R_{\mathrm{ns}})>\rho,
\end{array}\right.
\end{equation}
where $d$ is the distance of the fireball to the center of the magnetar and $\Delta A_1$ and $\Delta A_2$ are the areas of a portion of circles cut by a line segment, which are given by the angles subtended by the line segment as
\begin{eqnarray}
\Delta A_1&=&\frac{1}{2}l^2\left( \theta_1-\sin \theta_1 \right),\ \quad \theta_1=2\arccos \left( \frac{l^2+\rho ^2-R_{\mathrm{ns}}^2}{2\rho l} \right), \\
\Delta A_2&=&\frac{1}{2}R_{\mathrm{ns}}^2\left( \theta_2-\sin \theta_2 \right),\quad \theta_2=2\arccos \left( \frac{R_{\mathrm{ns}}^2+\rho ^2-l^2}{2\rho R_{\mathrm{ns}}} \right).
\end{eqnarray}\\
Therefore, it is straightforward to define the eclipse factor as
\begin{equation}
 f_e=\frac{A_{\mathrm{obs}}}{\pi l^2}.
\end{equation}
For the selected bursts, we assume their emission flux to be constant except during the fast-rising and decay phases. Specifically, we write the intrinsic light curve of the burst as follows:
\begin{equation}
    F_{\mathrm{int}} =F_{\mathrm{p}}\times \Theta \left( t-t_1 \right) \Theta \left( t_2-t \right), 
\end{equation}
where $t_1$ and $t_2$ are the starting and ending times of the burst plateau, $F_{\mathrm{p}}$ is the peak flux of the fireball emission without eclipse, and $\Theta(t)$ is the step function.
Therefore, the observed flux is simply given by
\begin{equation}
    F_{\mathrm{obs}}=f_e\times F_{\mathrm{int}}.
\end{equation}
Then, fitting the observed light curves of the eclipsed bursts can constrain the position and size of the emitting fireball.

\begin{figure*}[htbp]
\centering
\subfloat[Burst B: 2020-04-27 18:32:41.640]{\label{subfig:ecl_plateau_B}\includegraphics[width=0.5\textwidth]{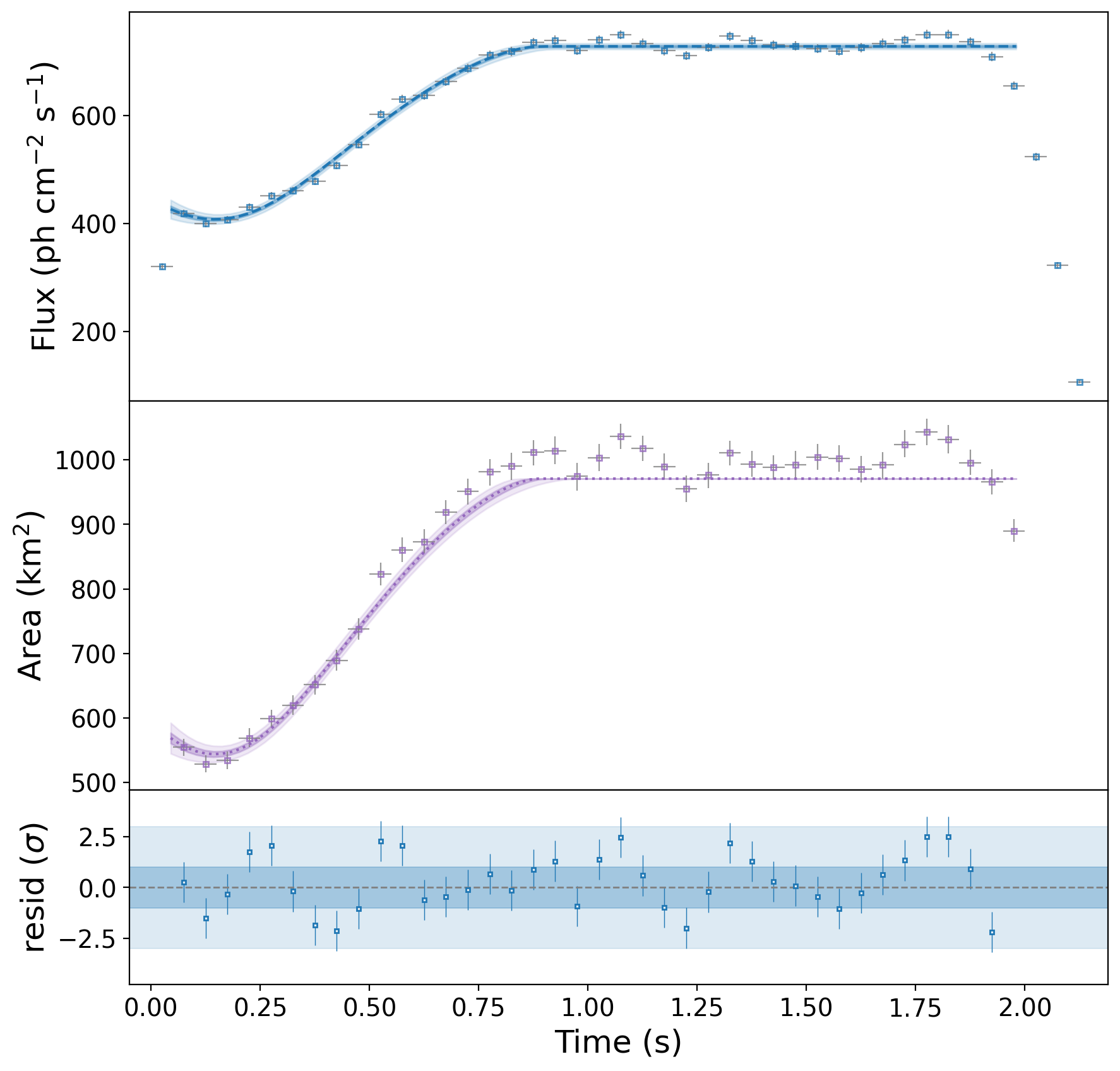}}
\hfill
\subfloat[Burst C: 2020-04-27 20:15:20.770]{\label{subfig:ecl_plateau_C}\includegraphics[width=0.5\textwidth]{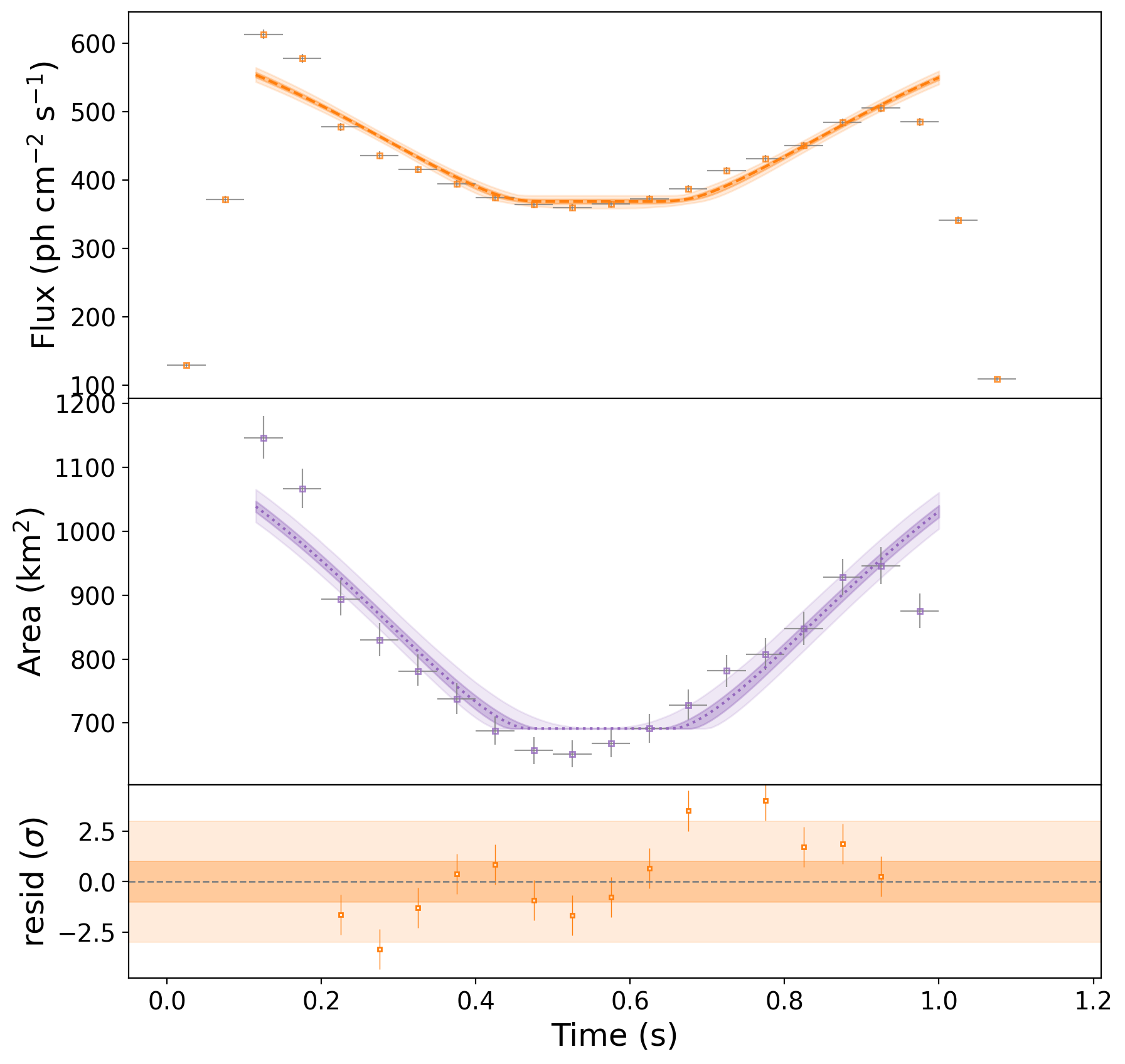}}
\hfill
\subfloat[Burst D: 2021-12-24 03:42:34.341]{\label{subfig:ecl_plateau_D}\includegraphics[width=0.5\textwidth]{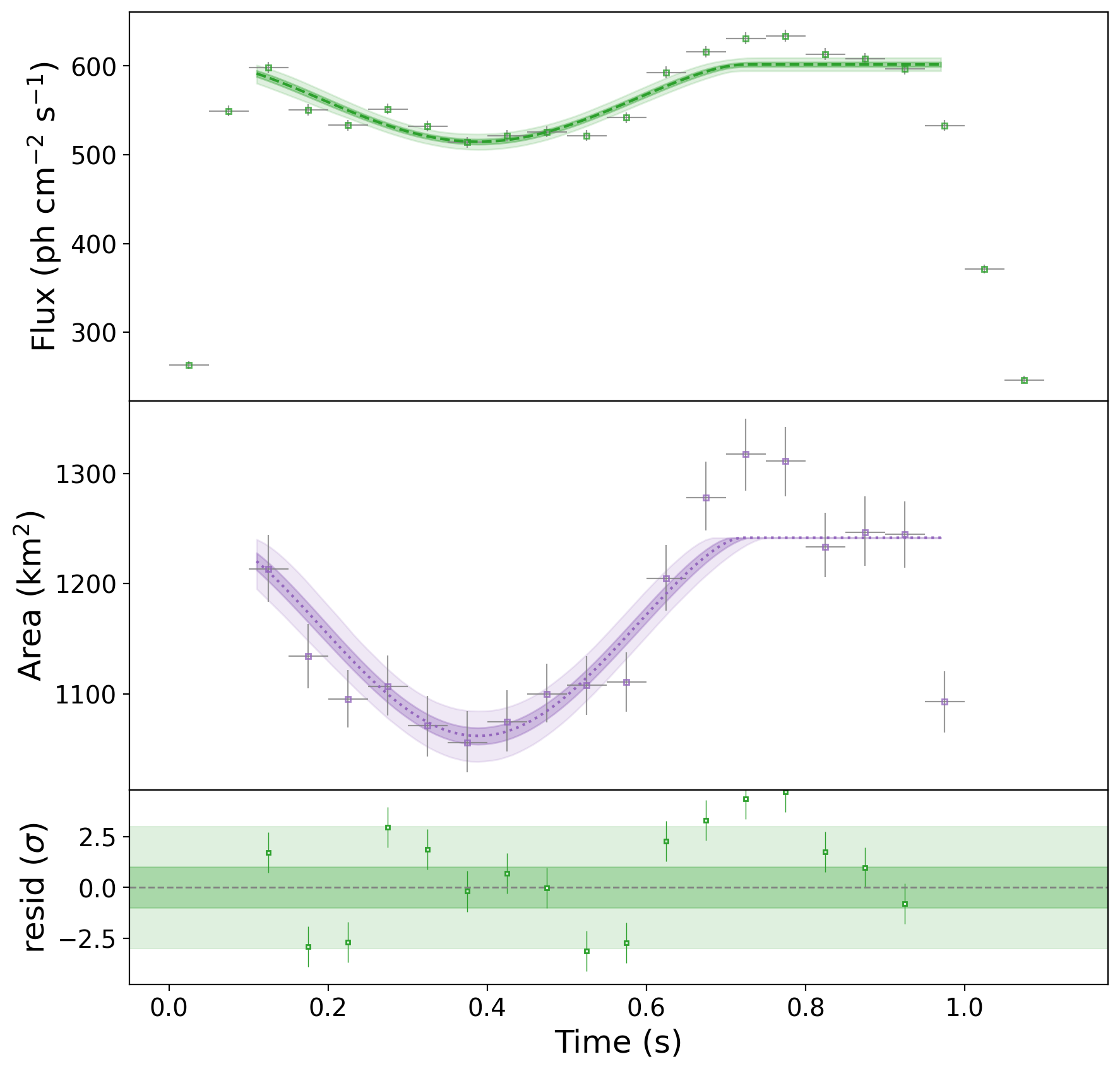}}
\hfill
\subfloat[Burst E: 2022-01-14 20:21:05.500]{\label{subfig:ecl_plateau_E}\includegraphics[width=0.5\textwidth]{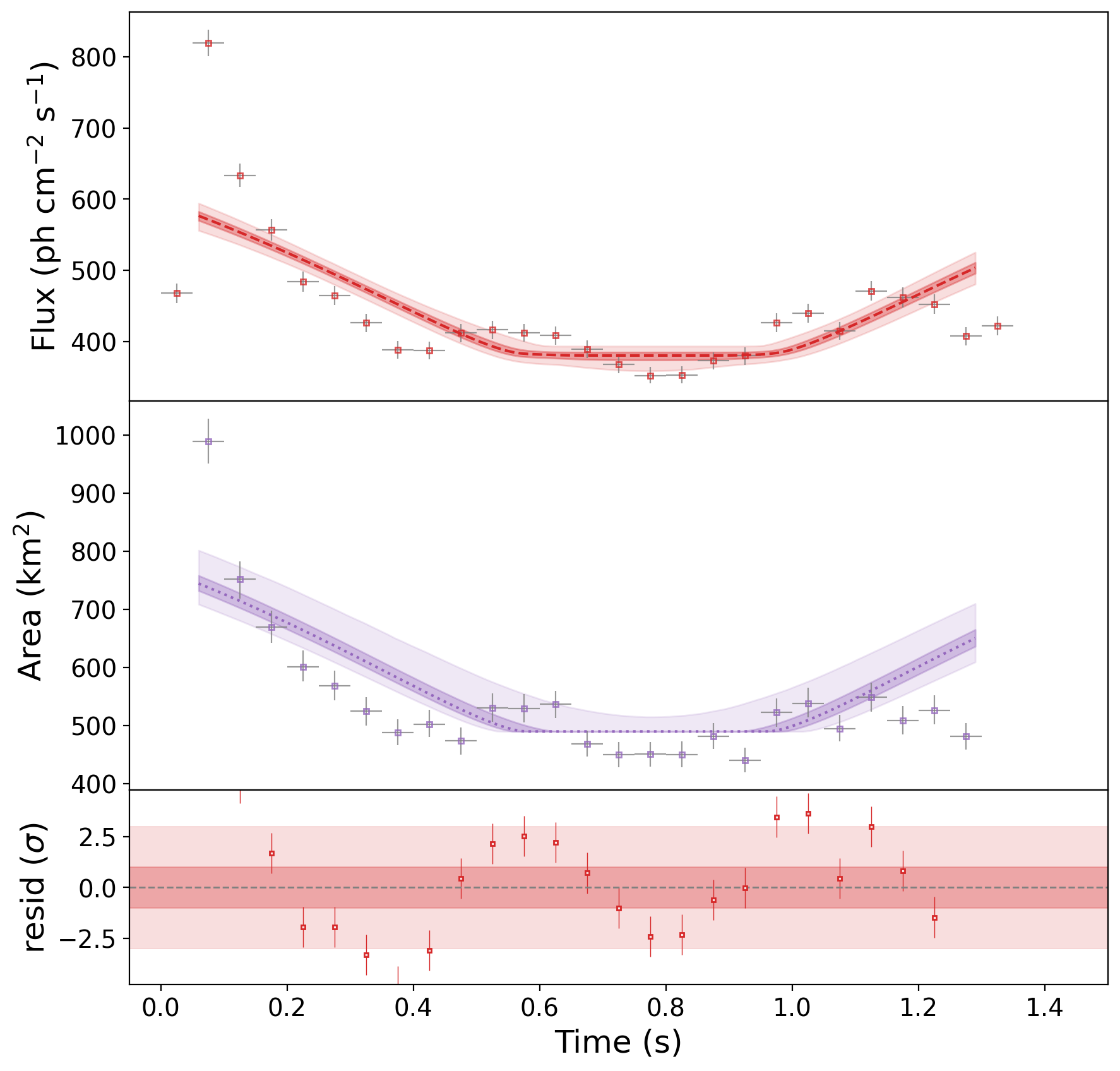}}
\caption{The temporal evolution behaviors of the eclipsed bursts during the plateau phase. The top panels present the evolution of the emission flux, where the dashed lines provide the fittings of these light curves with the eclipse model. In the middle panels, we plot the model-predicted evolutions of the BB emission areas (dotted line), in comparison with the data derived from the fittings of the time-resolved spectra. The residuals of the light-curve fittings are shown in the bottom panels, where the shadings represent $1\sigma$ and $3\sigma$ confidence levels.}\label{fig:lc_each_eclipsed}
\end{figure*}

\begin{figure*}[htbp]
\centering
\subfloat[Burst B]{\includegraphics[width=0.5\textwidth]{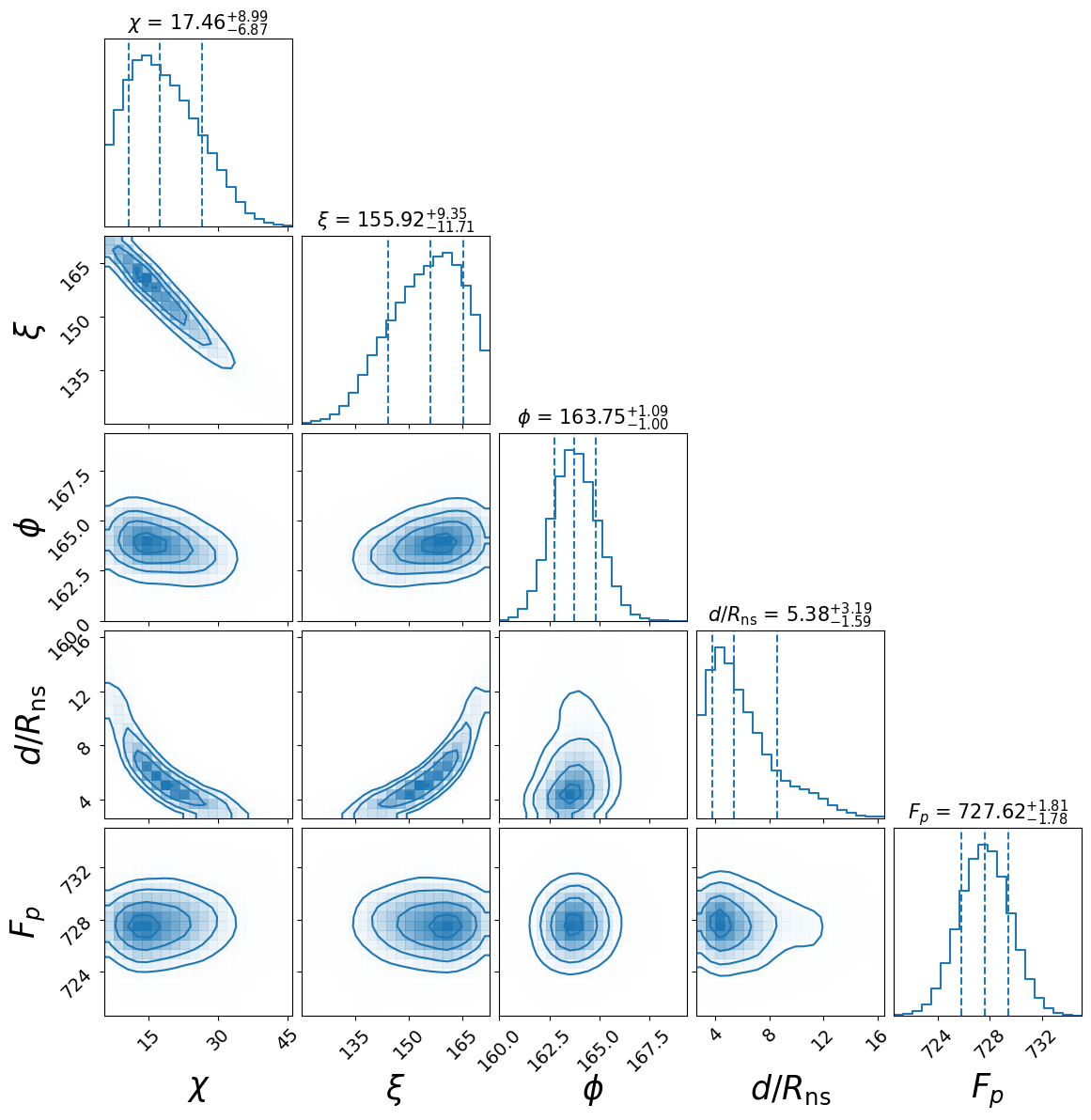}}
\hfill
\subfloat[Burst C]{\includegraphics[width=0.5\textwidth]{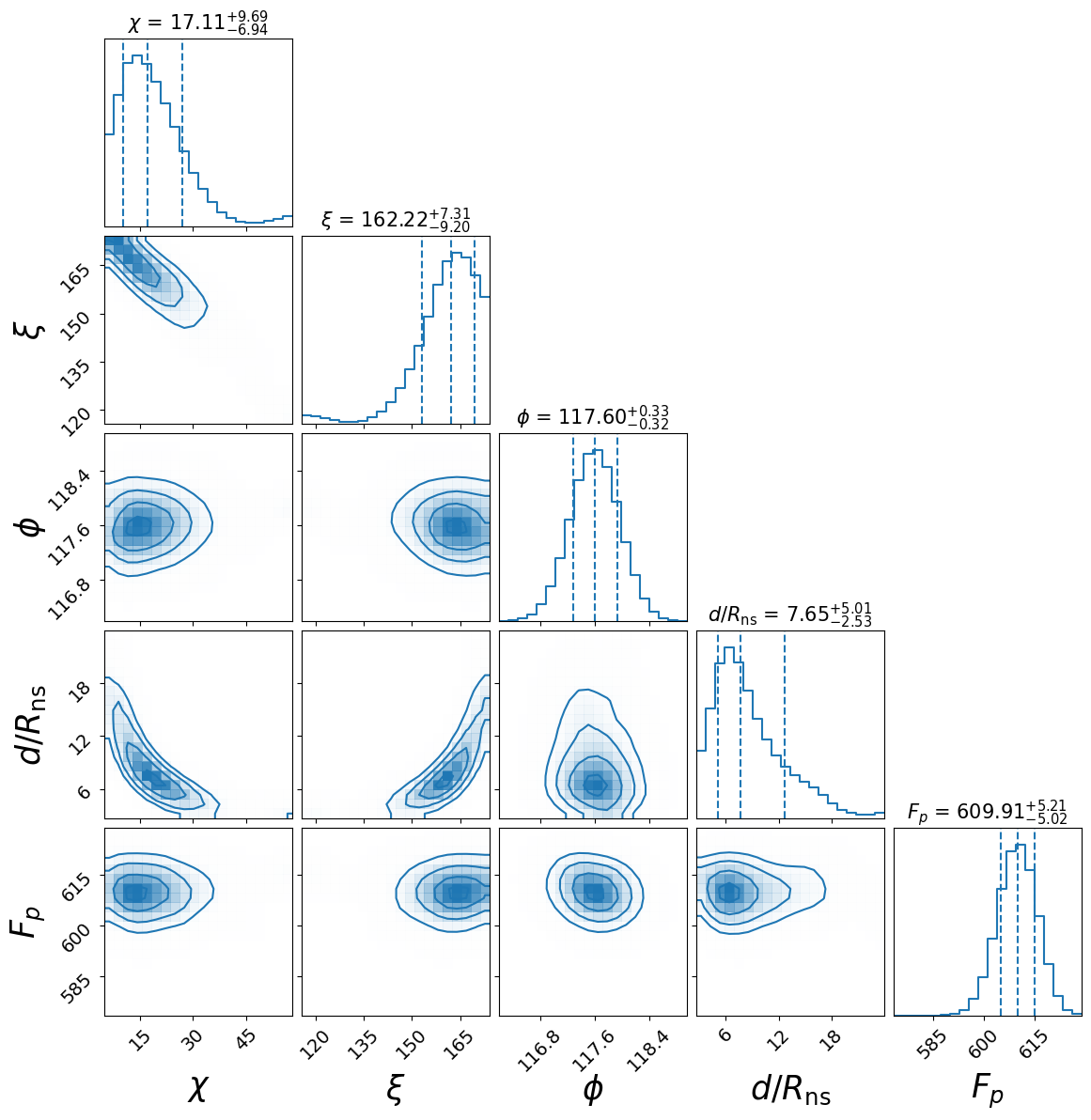}}
\hfill
\subfloat[Burst D]{\includegraphics[width=0.5\textwidth]{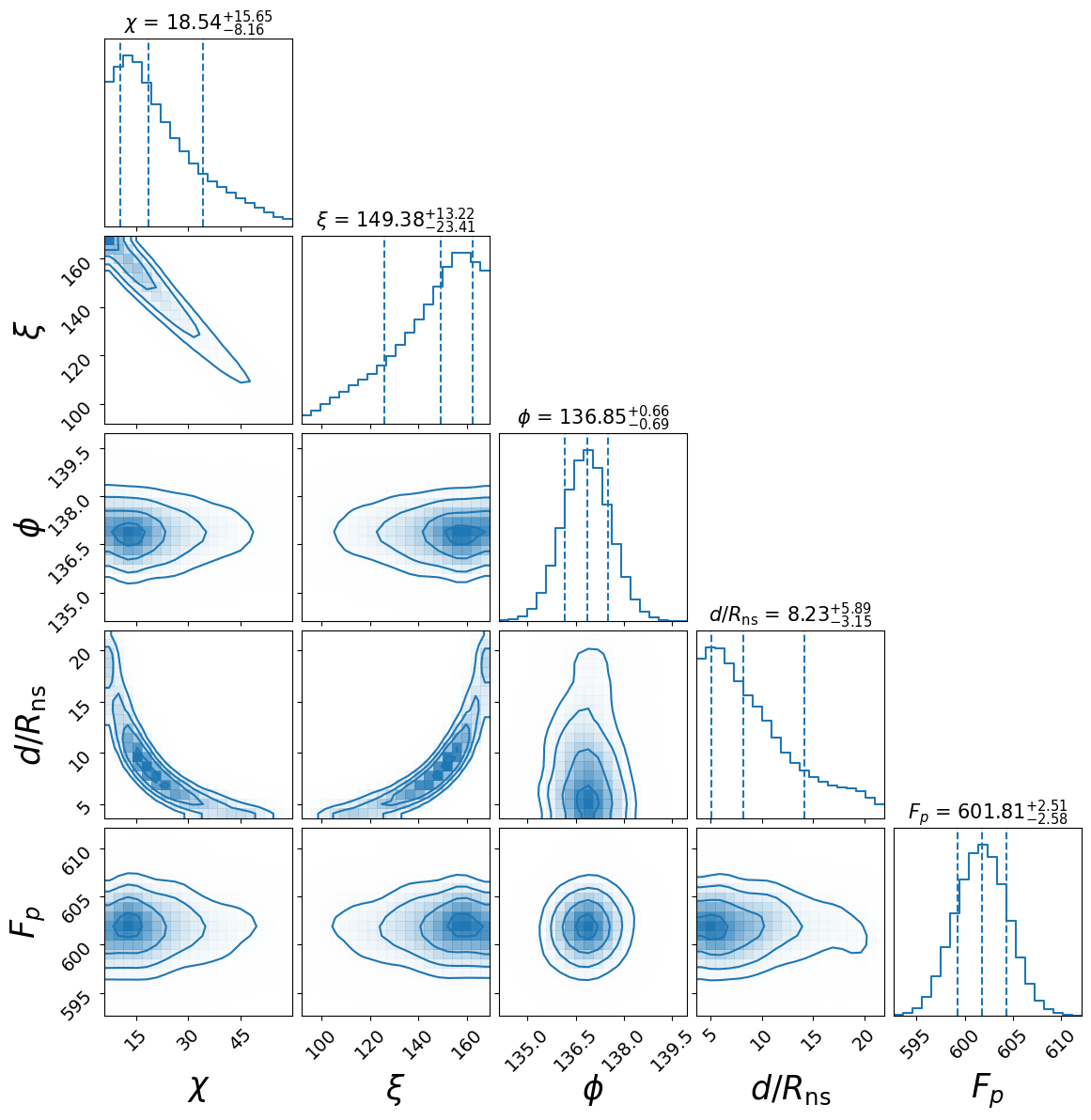}}
\hfill
\subfloat[Burst E]{\includegraphics[width=0.5\textwidth]{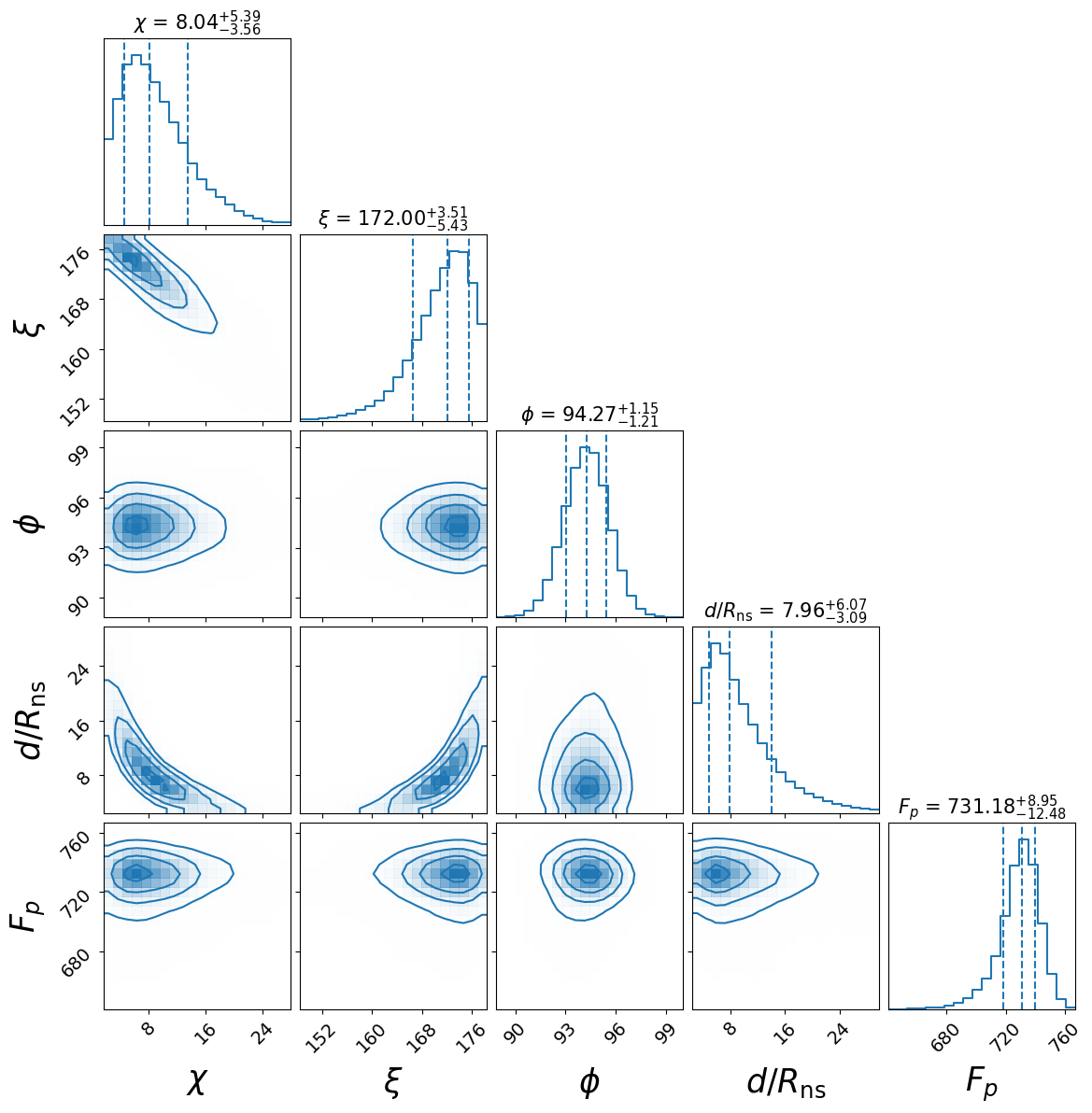}}
\caption{Posteriors of the light-curve fittings for Bursts B--E with the eclipse model. Medians and $1\ \sigma$ ranges of the parameters are labeled.}
\label{fig:eclipse_pos}
\end{figure*}

\begin{figure}[htbp]
\centering
\includegraphics[width=0.45\textwidth]{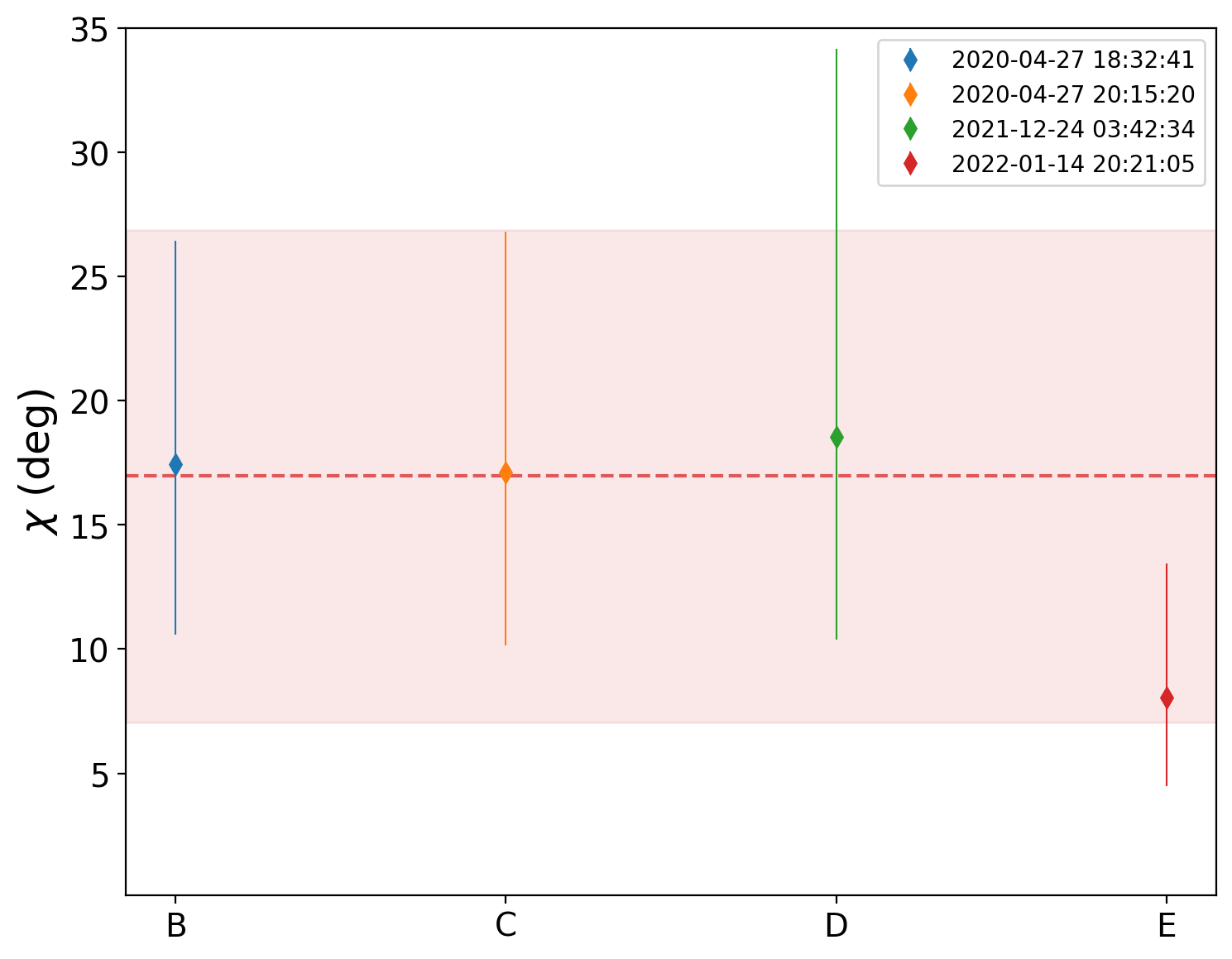}
\caption{The inferred viewing angle $\chi$ of the line of sight relative to the magnetar's spin axis. The dashed line and shaded band show the mean and standard deviation, respectively, of the four independent measurements.}
\label{fig:dist_chi}
\end{figure}

With this model, we fit the light curves of the four eclipsed bursts, assuming an intrinsically steady fireball, as shown by the dashed lines in Figure \ref{fig:lc_each_eclipsed}, with the corresponding posterior distributions shown in Figure \ref{fig:eclipse_pos}.
The corresponding parameter values with errors are listed in Table \ref{tab:elc_results}. In the calculations, the intrinsic radii of the fireballs are taken from the spectral fits.
The viewing angles derived from the four eclipsed bursts are found to be all concentrated around a common value
\footnote{The common value is estimated from the mean of individual posteriors and includes uncertainties,
\begin{equation}
\bar{\chi} = \frac{1}{4} \sum_{i=B}^{E} \bar{\chi}_i, \quad
\sigma^2 = \frac{1}{4} \sum_{i=B}^{E} \left[ \sigma^2_i + (\bar{\chi}_i - \bar{\chi})^2 \right],
\end{equation}
where $\bar{\chi}_{i=\mathrm{B,...E}}$ and $\sigma_{i}^2$ is the posterior mean and variance of the $i$th burst, respectively.}
of $\bar{\chi} = 17^\circ \pm 10^\circ$, as shown in Figure \ref{fig:dist_chi}, indicating the self-consistency of our modeling. 
We stress that the pulsar (neutron star) viewing angle can be determined through this novel eclipse-based method, in contrast to the conventional rotating vector model that relies on radio polarization measurements. The derived values of the parameter $\xi$ are relatively large, because they are only for those fireballs located at the opposite side of the observer that would be eclipsed by the magnetar itself. Our results further show that $d\gg R_{\mathrm{ns}}$ and $d\gg l$, which indicate that the emitted fireballs of these bursts are far away from but do not adhere to the magnetar's surface. This is consistent with the implication of the CRSFs. 
Finally, the distance $d$ is much smaller than the radius of the light cylinder as $R_{\rm LC}=cP/2\pi=1.5\times10^5$ km, suggesting that the fireballs are located deep in the magnetosphere. 

It is necessary to note that in our corotating spherical fireball model we focus on the intermediate plateau phase of the burst, during which the emitting region approximately corotates with the magnetosphere. Physically, the fireball is expected to undergo rapid expansion and subsequent cooling and contraction, as in classical trapped fireball scenarios, but these processes correspond to the short rise and decay phases and occur on timescales much shorter than the plateau duration. We therefore approximate the fireball as a quasi-steady, uniform emitter during this phase, which allows us to investigate how global magnetospheric rotation and geometrical viewing angles imprint on the observed flux modulation without modeling the detailed microphysical evolution.
Furthermore, we do not account for gravitational lensing in our model, which can slightly increase the brightness of a distant light source if the light grazes the magnetar's surface. However, given that the fireball is comparable to or several times larger than the magnetar in size, we should treat the fireball as an extended source rather than a point source. In such a case, since different differential elements on the source have different distances to the caustic, the magnification of the total brightness of the fireball should be averaged over the entire extended surface. As a result, the brightness variation arising from the distance change of the fireball to the caustic would be much smaller than that for a point source \citep{DallOsso2024ApJ}.

\begin{table*}
\centering
\caption{The observational data utilized and the eclipse model fitting parameters. The errors are given at the $1 \sigma$ confidence level. The symbols $l$, $d$, $\chi$, and $\xi$ are defined as illustrated in Figure \ref{fig:geometry}. $\phi$ is the phase corresponding to the starting time, $R_{\mathrm{ns}}$ is the radius of the magnetar, and $F_{\mathrm{p}}$ is the peak flux of the fireball emission without eclipse.}
\label{tab:elc_results}
{\footnotesize
\begin{tabular}{l|ccccc}
\toprule
Labeled Burst & A & B & C & D & E \\
\hline
Start time 
&2016-06-26 
&2020-04-27 
&2020-04-27 
&2021-12-24 
&2022-01-14 \\
&13:54:30.720 
&18:32:41.640
&20:15:20.770
&03:42:34.341
&20:21:05.500 \\
\hline
Instrument & GBM & GBM & GBM & GBM & GECAM-B \\
Detectors & n6, n9 & n6, n9 & n2, n9, na & n1, n3, n5 & g1, g6, g7 \\
Energy (keV) & 8-80 & 8-80 & 8-80 & 8-80 & 20-80 \\
Duration (s) & 0.84 & 2.15 & 1.09 & 1.3 & 1.5 \\
Fireball radius (km) & $17.44_{-0.23}^{+0.45}$ & $17.58_{-4.61}^{+0.63}$ & $19.08_{-1.67}^{+2.05}$ & $19.88_{-1.54}^{+0.59}$ & $17.32_{-5.50}^{+0.42}$ \\
\hline
Model Parameters  & Priors & \multicolumn{4}{c}{Posteriors} \\
\hline
$\chi$ (deg) & [0--180] & $17.46_{-6.87}^{+8.99}$ & $17.11_{-8.94}^{+9.69}$ & $18.54_{-8.16}^{+15.65}$ & $8.04_{-3.56}^{+5.39}$ \\
$\xi$ (deg) &  [0--180] & $155.92_{-11.71}^{+9.35}$ & $162.22_{-9.20}^{+7.31}$ & $149.38_{-23.41}^{+13.22}$ & $172.00_{-5.43}^{+3.50}$ \\
$\phi$ (deg) &  [0--360] & $343.75_{-1.00}^{+1.09}$ & $297.60_{-0.32}^{+0.33}$ & $316.85_{-0.69}^{+0.66}$ & $274.27_{-1.21}^{+1.15}$ \\
$d/R_{\mathrm{ns}}$ & [1--50] & $5.38_{-1.58}^{+3.19}$ & $7.65_{-2.53}^{+5.01}$ & $8.23_{-3.15}^{+5.89}$ & $7.96_{-3.09}^{+6.07}$ \\
$F_{\mathrm{p}}$ (photons cm$^{-2}$ s$^{-1}$) &  
[1--1000] & $727.62_{-1.78}^{+1.81}$ & $609.91_{-5.02}^{+5.21}$ & $601.81_{-2.58}^{+2.51}$ & $731.18_{-12.48}^{+8.95}$ \\
\hline
$\chi^2$/dof & \ldots & 14.78/34 & 27.25/13 & 15.94/13 & 75.76/19 \\
\botrule
\end{tabular}}
\end{table*}

\section{Discussion}\label{sec:dicuss}
The distance measurements of the fireballs to the magnetar surface provide a very stringent constraint on the trigger mechanism of the bursts. Generally speaking, highly twisted magnetic fields in the magnetosphere of magnetars are very likely to extend from the stellar interior, where the toroidal component could dominate. In other words, the energy release of magnetars is ultimately a result of the diffusion of magnetic helicity from the interior to the exterior. The connection between the internal and external fields exerts vast stress on the solid crust of magnetars. Then, along with the spin-down of magnetars, the shear stress is gradually accumulated to finally exceed the yield limit of the crust, leading to plastic deformation of the crust and triggering a starquake. In principle, the released strain energy can heat the deforming crustal region, leading to the formation of a hot spot and even a fireball trapped by local magnetic field loops near the surface.

However, such a picture is now disfavored for these eclipsed plateau bursts, as their distances are approximately five times the stellar radius. Instead, the fireballs are inferred to be suspended in the magnetosphere, suggesting that they originate from local magnetic reconnection of magnetic ropes or knots. The magnetic ropes/knots may form as the cumulative product of long-term crustal deformation. Thus, the occurrence of such catastrophic magnetic reconnection is not necessarily linked to an individual starquake, even though the energy released ultimately derives from the cumulative effect of numerous previous starquakes.
It is worth pointing out that our conclusions are only applicable to the selected bursts with intrinsic plateau emission and observed eclipsing feature. So far, our analysis does not rule out the starquake scenario that also forms a fireball.

Furthermore, we can envision the following physical scenario: As a magnetic reconnection occurs in the magnetosphere, an energy of about $E_{\rm iso}\sim10^{41}$ erg can be released suddenly. Then, a great number of electron and positron pairs can be excited in the reconnection zone and flow into a larger-scale region, forming a magnetically trapped fireball.
The number density of the fireball can be given by \citep{Landau1980,Thompson1995MNRAS}
\begin{equation}
n_{\pm}={1\over 2} \left({2 m_{\rm e}  k_{\rm B}  T_{\rm in}\over \pi  \hbar^2}\right)^{3/2} \exp\left(-{m_{\rm e}  c^2\over k_{\rm B}  T_{\rm in}}\right),\label{npm}
\end{equation}
where $e$ and $m_{\mathrm{e}}$ are the charge and mass of electrons, respectively, and $c$ is the speed of light.
The internal temperature of the fireball $T_{\rm in}$ can be roughly estimated to be $k_{\rm B}T_{\rm in}=k_{\rm B}({3E_{\rm iso}/ 4\pi l^3a})^{1/4}=68(E_{\rm iso}/ 10^{41}{\rm erg})^{1/4}(l/ 20{\rm km})^{-3/4} \rm keV$ with $a$ being the radiation constant. The fireball expansion could be quickly suppressed if the pressure of the surrounding large-scale magnetic fields can finally exceed the thermal pressure of the fireball, which requires 
$B\gtrsim 1.6\times10^{11}\: ( {l}/{20\:\mathrm{km}} )^{-3/2} ( {E_\mathrm{iso}}/{10^{41}\: \mathrm{erg}} )^{1/2}\: \mathrm{G}$ \citep{Thompson1995MNRAS}.
By using a simple dipolar field configuration, the field strength at the distance $d$ can be estimated as $B(d)\sim 3.0\times10^{12} ({d}/{50\,\mathrm{km}} )^{-3}\,\mathrm{G}$, which is obviously strong enough to stop the fireball expansion, where a reference stellar radius $R_{\mathrm{ns}}=12$ km is adopted and the surface field strength is taken as $B_\mathrm{p}=2.2\times10^{14}\: \mathrm{G}$ \citep{Israel2016MNRAS}. Therefore, the fireball can finally be trapped by the magnetic fields and suspended in the magnetosphere for a time interval. 

The rationality of a suspended fireball may also be understood from the perspective of interpreting the key properties of the CRSF absorption, including its central energy, line width, and absorption depth: (i) With the magnetic field strength at a reference distance of 50 km, we can determine the position of the first harmonic of CRSFs in the spectra of the bursts as 
$E_\mathrm{cyc}=\hbar {eB/ m_{\rm e}c}=35.3 \left({d/ 50\ {\rm km}}\right)^{-3}\rm keV$, which is valid for the case of $B\ll B_{\mathrm{c}}=4.4\times10^{13}\,\mathrm{G}$ with $B_{\mathrm{c}}$ being the critical value for Landau quantization. Although very simple, this estimate of $E_\mathrm{cyc}$ can be very consistent with the values obtained from the spectral fittings, further indicating the self-consistency of the model. However, strictly speaking, some inconsistencies still exist, since the fireball distances inferred from the light-curve fittings (see Table \ref{tab:elc_results}) are actually somewhat different from the reference value. Here the crux of the problem may lie in the complication of the magnetic field configuration of the magnetar, which distinguishes itself from the traditional dipolar scenario and complicates the determination of both $B(d)$ and $E_{\rm cyc}$ at a given distance. (ii) It is expected that the widths of CRSFs are at least larger than that determined by the Doppler effect as \citep{Meszaros1985,Harding1991ApJ,Nishimura2003,Schonherr2007A&A} 
$\sigma=({k_{\rm B}T/ m_{\rm e}c^2})^{1/2}E_{\rm cyc}|\cos \theta|
=4.9{\rm keV}\left({k_{\rm B}T/{10\rm keV}}\right)^{1/2}\left({E_{\rm cyc}/35\rm keV}\right)|\cos \theta| $, where $\theta$ is the angle between the photon direction and the magnetic field vector. This value is smaller than, but of the same order of magnitude as, the CRSF widths derived from the spectral fittings. However, since the fireball is actually distributed within a spatial range of a few tens of kilometers, the magnetic field strength in the vicinity of the fireball should at least vary on an amplitude of $[(d+l)/(d-l)]^3$. Given this significant variation in the field strength, it would be not surprising to observe a CRSF line a few times broader than the Doppler broadening. (iii) In the outer region adjacent to the fireball surface where the CRSF could be formed, the column density of electrons and positrons can be estimated by $\mu n_{\rm GJ}\Delta$, where $\mu$ is the multiplicity, $n_{\rm GJ}=\Omega B/(2\pi c e)$ is the Goldreich--Julian density, and $\Delta$ is the thickness of this line formation region. Then, following \cite{Wang1993ApJ}, the optical depth of the CRSF absorption can be expressed as
$\tau_\mathrm{CRSF}=0.3\left({\mu/ 10^{3}}\right)\left({\Delta/ 5~\rm km}\right)\left({k_{\rm B}T/ \rm 10~keV}\right)^{-1/2}$, 
which could provide a basic explanation for the optical depths derived from the spectral fittings.

The dynamical evolution of a fireball is determined by the competition between the internal thermal pressure and external magnetic pressure. Supposing that a burst is triggered impulsively, a plateau emission would occur during a pressure balance. Then, as the majority of the internal energy of the fireball is lost, the balance would be disrupted, and thus the plateau emission ceases. Therefore, we can express the duration of the plateau emission in terms of the diffusion timescale of thermal photons in the fireball as
\begin{equation} 
t_{\rm diff}\sim{l^2\over c \lambda }=3.3\times10^4\left({k_{\rm B}T_{\rm in}\over 1~\rm keV}\right)^{3/2}\exp\left(-{511{\rm keV}\over k_{\rm B}T_{\rm in}}\right) ~{\rm s},\label{eq:tdiff}
\end{equation}
where $\lambda=1/(\sigma_{\rm T}n_{\pm})$ is the mean free path of the thermal photons, where $\sigma_{\rm T}$ is the Thomson cross section and $l=20\,\mathrm{km}$ is adopted for giving the numerical coefficient. However, this timescale is very sensitive to the determination of the internal temperature of the fireball because of the presence of the exponential term. Specifically, by solving equation $t_{\rm diff}\sim 1$ s, we can obtain $k_{\rm B}T_{\rm in}=32.7$ keV for $l=20\,\mathrm{km}$, which is lower than that inferred from the average internal energy density. Here it should be noted that the actual temperature within the fireball must exhibit a decreasing distribution from the inside out, which is determined by thermal transfer within the fireball. Given this, it is not surprising to obtain different temperatures from different estimation methods, not to mention that these approaches are rather crude. Then, undoubtedly, a precise calculation of the burst duration requires a detailed characterization of the thermal diffusion.

In addition, more complexity can arise from the unclear trigger mechanism of the bursts. Given the complication of locally twisted magnetic fields, magnetic reconnection is likely to occur repeatedly and with complicated dynamical behavior rather than impulsively as supposed above. In this case, the pressure balance of the fireball cannot remain steady, and thus an evolving and even multipeaked light curve rather than a plateau would be obtained. Such complicated bursts can be easily found from the longest bursts of SGR J1935+2154, with a larger number than plateau bursts. Since we cannot know the dynamical evolution of the fireballs of these complicated bursts, it is actually impossible to identify eclipse candidates from them, even though their spectral properties are not obviously different from the plateau ones. 

Furthermore, it is indicated that plateau bursts are actually very rare, which seems sound because of their strict requirement on the burst trigger mechanism. In any case, these facts make it difficult for us to verify the eclipse model from the perspective of number statistics.

\section{Summary}\label{sec:summary}
In this paper, we identified one representative plateau-like burst among the long bursts of SGR J1935+2154 detected by GECAM and Fermi/GBM, and found four candidate bursts showing eclipse-like features. 
We established the corotating fireball geometric eclipse model and fitted the light curves of four eclipse bursts. We derived the viewing angle related to the spin axis with $\chi \approx 17^{\circ}$, and the fireball is located at several neutron star radii away from the magnetar. This distance indicates that the fireball is suspended in the magnetosphere, where the magnetic field strength is estimated to be $B \sim 3\times10^{12}~\mathrm{G}$. Moreover, we find CRSF features in the spectra of X-ray bursts, which also implies a magnetic field strength similar to that inferred from the light-curve distances. 
These results not only provide smoking-gun evidence of the corotating fireball, which accounts for the energetic X-ray emission during the magnetar activity, but also support an origin of the fireball from magnetic reconnection in the magnetosphere rather than directly from a crustal starquake. We also note that this new eclipse-based method provides a novel tool to constrain the viewing angle and could be applied to other magnetar bursts.

\begin{acknowledgments}
This work is supported by the National Key R\&D Program of China (2021YFA0718500), 
the National SKA Program of China (2020SKA0120300),
the National Natural Science Foundation of China (grant Nos. 12393811, 12393812, and
12273042
),
the Strategic Priority Research Program, the Chinese Academy of Sciences (Grant No. 
XDB0550300
) and the China's Space Origins Exploration Program. 
The GECAM (HuaiRou-1) mission is supported by the Strategic Priority Research Program on Space Science of the Chinese Academy of Sciences. We appreciate the public data and software of Fermi/GBM.
\end{acknowledgments}

\facilities{GECAM, Fermi/GBM}


\appendix
\restartappendixnumbering

\section{Data Reduction}\label{appxsec:data_reduct}
GECAM \citep[X.][]{LiWen2021,Xiao2022mnras} is originally a constellation of two small X-ray and gamma-ray all-sky observatories funded by the Chinese Academy of Sciences. These two microsatellites, denoted as GECAM-A and GECAM-B, were launched in 2020 December. Each GECAM satellite is equipped with 25 gamma-ray detectors (GRDs) and 8 charged particle detectors (CPDs). Each GRD operates in two readout channels, high gain (HG; 20--300 keV) and low gain (LG; 300 keV--5 MeV), which are independent in terms of data processing, transmission, and dead time. The time resolution of GRDs is 0.1 $\mu \rm s$. We select detectors with detector-to-source angles of $\leqslant50^\circ$ in our analysis. Given the dominated emission energy range of magnetar bursts ($\lesssim$100 keV), we focus on using the HG channel for data analysis. The GECAM data reduction tool is \texttt{GECAMTools} \citep{GECAMTools2024}.
Considering the lanthanum absorption edge at 38.89 keV for \(\mathrm{LaBr_{3}}\), as this feature potentially leads to systematic uncertainties in spectra, we have ignored the data in the range of 35--40 keV \citep{HeAn2023MNRAS}.

Fermi/GBM \citep[][]{Meegan2009apj} has a continuous broadband energy coverage that which consists of 12 NaI detectors (8--1000 keV) and 2 bismuth germanate detectors (0.2--40 MeV). GBM provides three kinds of science data: CSPEC with continuous high energy resolution (1024 ms and 128 energy channels), CTIME with continuous high time resolution (64 ms and 8 energy channels), and time-tagged photon event (TTE) with temporal resolution of $2\,\mu \mathrm{s}$ and energy resolution of 128 channels. We choose the detectors with good detector-to-source angles ($\leqslant50^\circ$), and hence the highest signal-to-noise ratio \cite{Collazzi2015apjs,Lin2020apj}. The data reduction tool used to generate light curves and spectral data is \texttt{GBM Data Tools} \citep{GbmDataTools}. Response files are generated by \texttt{GBM Response Generator v1.13} for spectral analysis.
Similar to GECAM, we excluded 30--40 keV data for NaI detectors to avoid the iodine K edge at 33.17 keV \citep{Bissaldi2009ExA}.

\section{Spectral Analysis}\label{appxsec:spec_analys}
\begin{figure*}[htbp]
\centering
\subfloat[Burst A]{\includegraphics[width=0.45\textwidth]{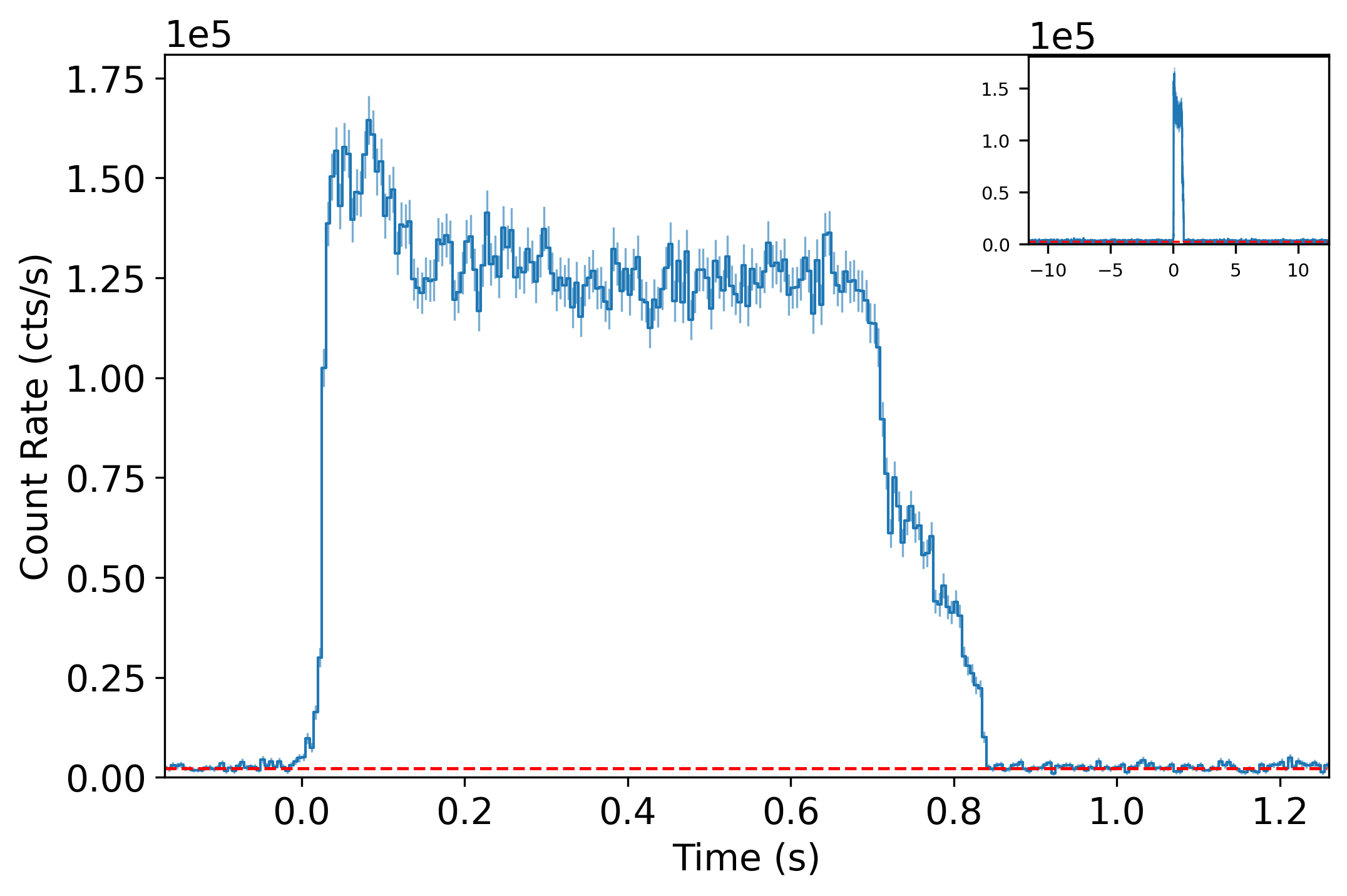}}
\hfill
\subfloat[Burst B]{\includegraphics[width=0.45\textwidth]{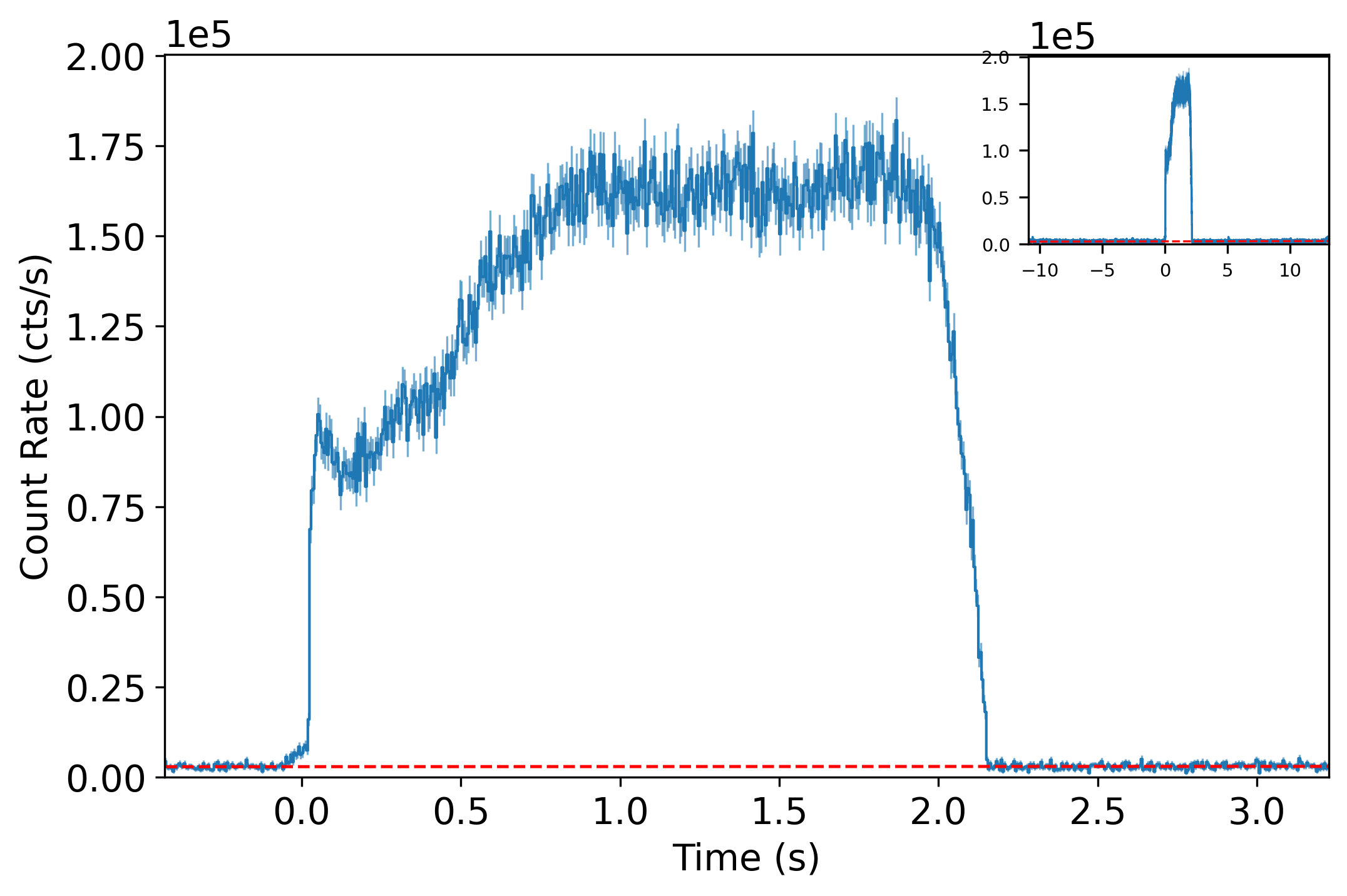}}
\hfill
\subfloat[Burst C]{\includegraphics[width=0.45\textwidth]{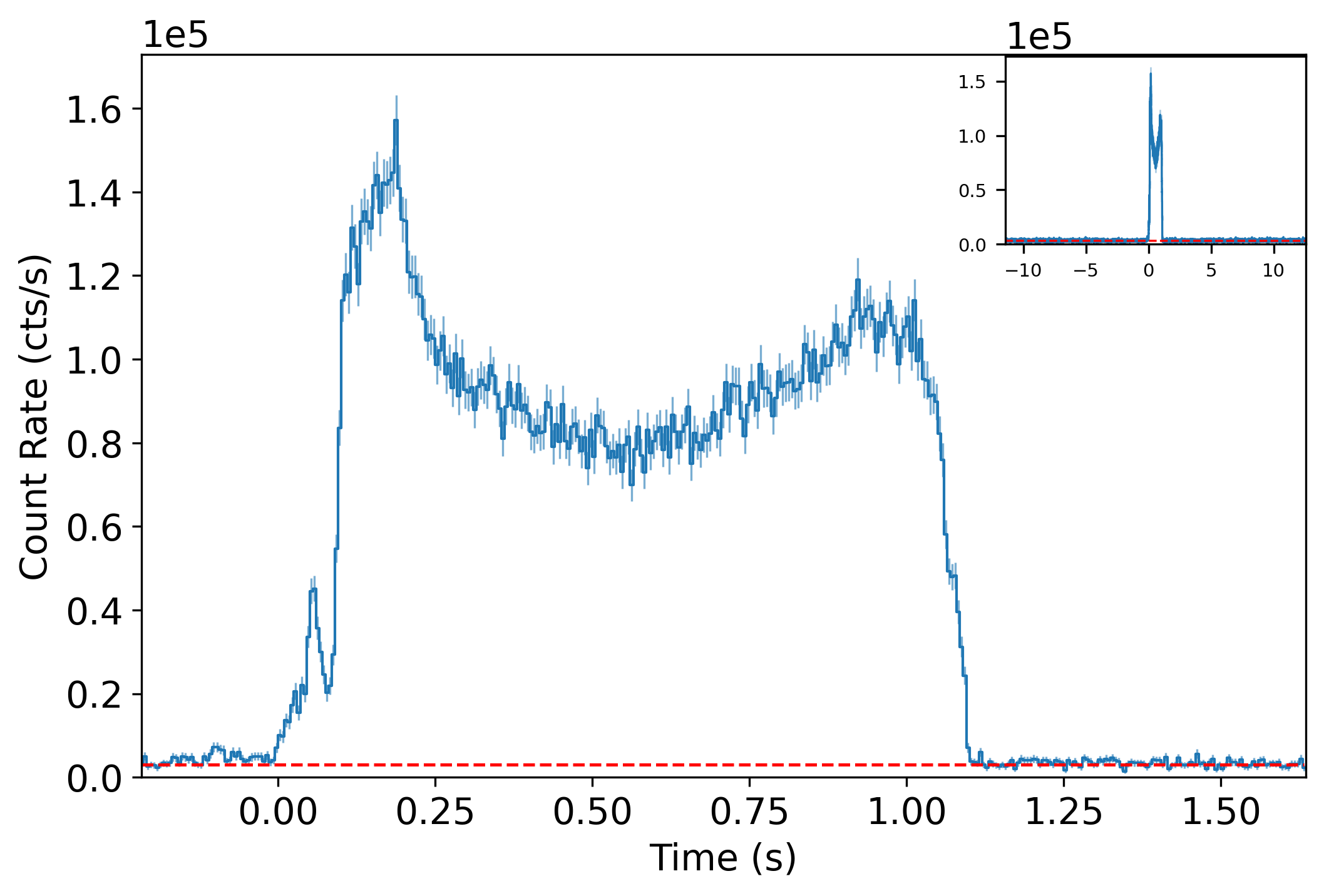}}
\hfill
\subfloat[Burst D]{\includegraphics[width=0.45\textwidth]{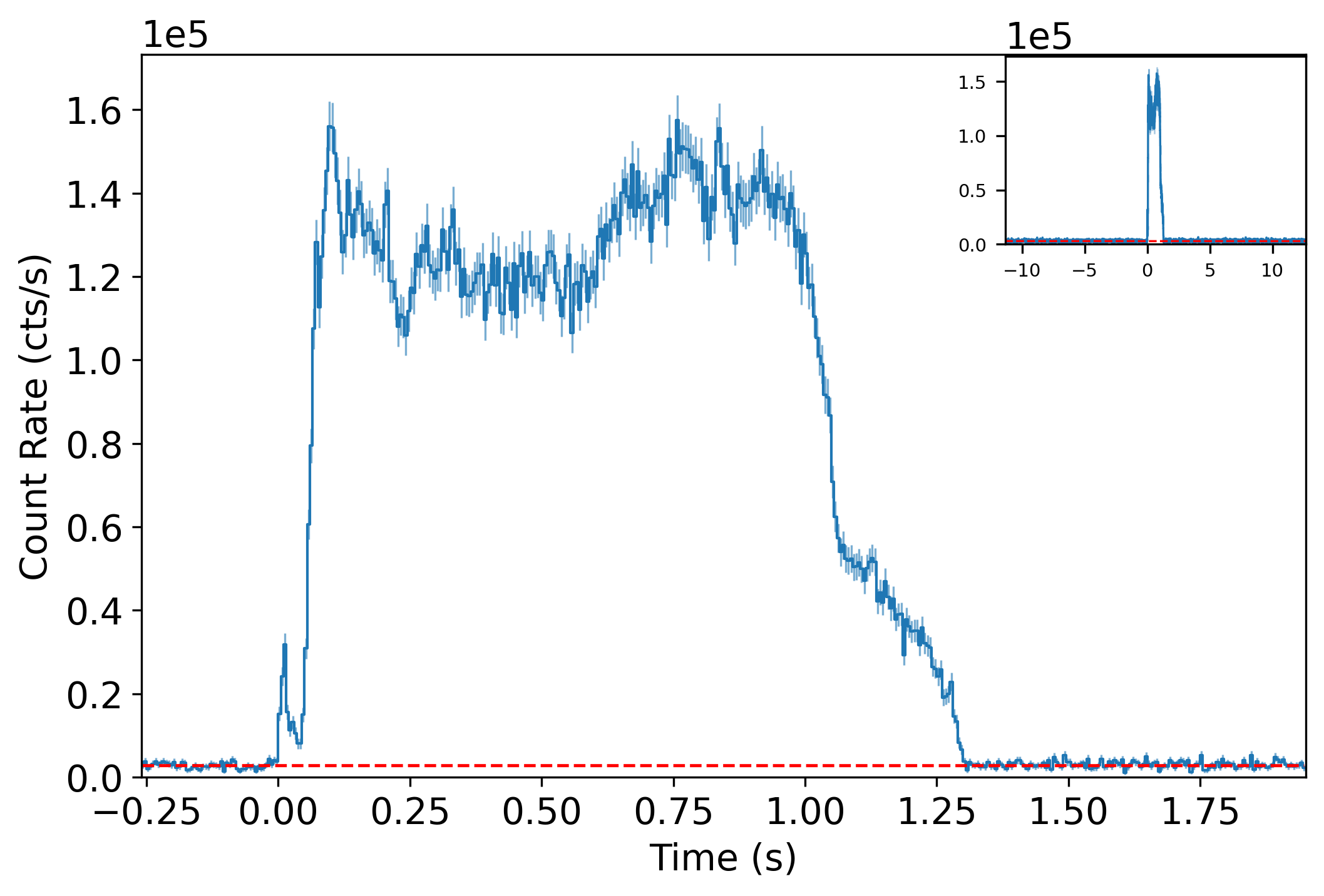}}
\hfill
\subfloat[Burst E]{\includegraphics[width=0.45\textwidth]{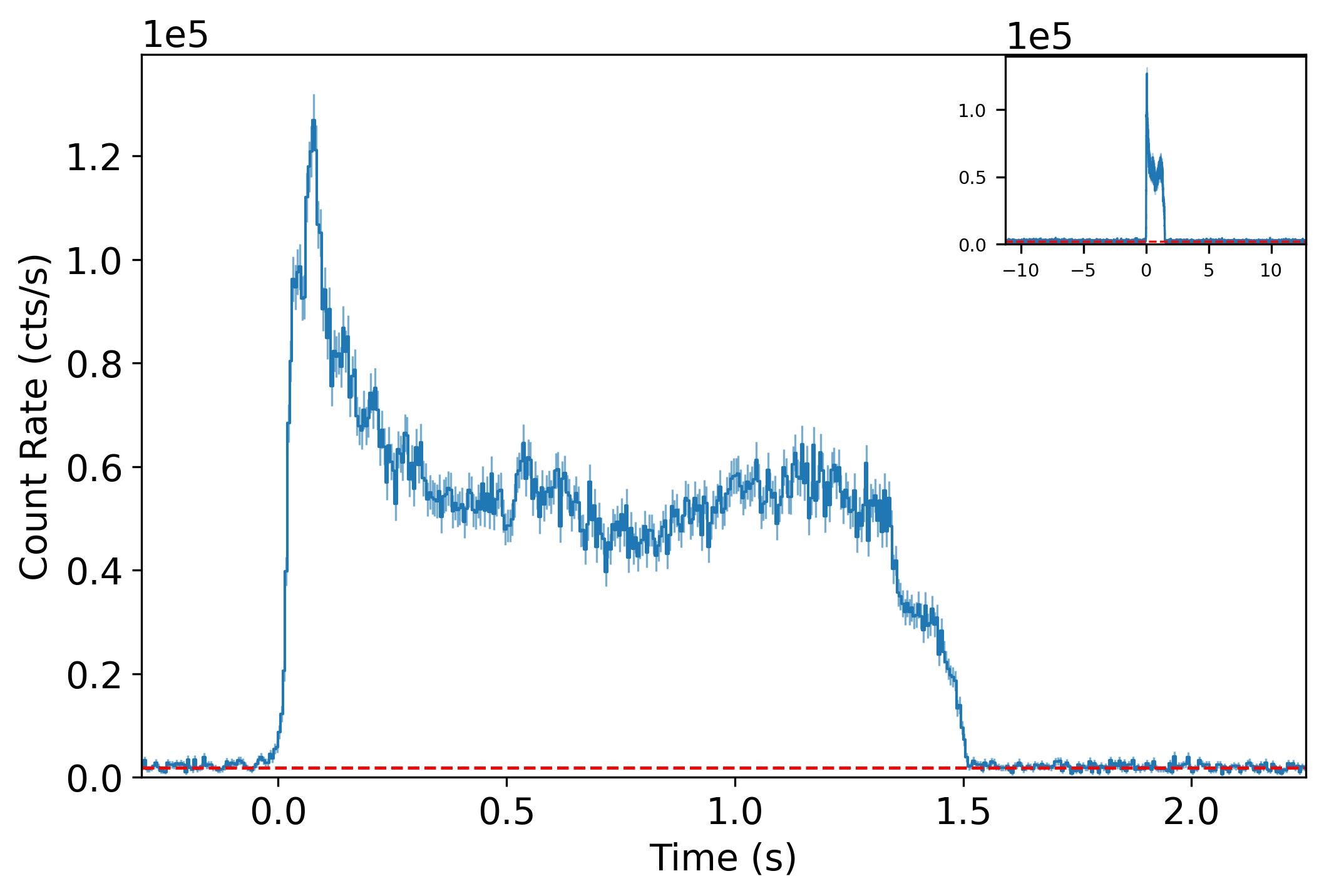}}
\caption{The raw light curves of Bursts A--E shown with a 5 ms time bin. The red lines represent the background. The insets show the light curves within a wider time range to illustrate the clean background level.}
\label{appxfig:raw_lc}
\end{figure*}

We process the GECAM event-by-event (EVT) data and GBM TTE data to generate spectral data for 50 ms time-resolved spectral analysis using a data reduction tool. Background spectra are generated from data events in pre- and post-burst time intervals (from $t_0-10\,\mathrm{s}$ to $t_0-5\,\mathrm{s}$, and from $t_0+5\,\mathrm{s}$ to $t_0+10\,\mathrm{s}$, where $t_0$ is the burst time) and fitted with a polynomial function to model the background during the burst. Subsequently, this background would be subtracted from the total spectrum in the spectral fitting process. 
The raw light curves of the five selected bursts, with a 5 ms time resolution and the estimated background level, are shown in Figure \ref{appxfig:raw_lc}. These light curves are very unlikely to result from the simple superposition of multiple pulses, although some nontrivial substructures may still exist.

The software \texttt{elisa} is employed for exploring and analyzing the spectral data \citep{ELISA}. 
In performing Bayesian fitting, the likelihood based on the $\chi^2$-statistic is used to compute the posterior distribution, and spectral channels are grouped to ensure a minimum of one count per channel.
Furthermore, to evaluate the performance of different spectral models, we used two commonly adopted information criteria: the AIC and the BIC \citep{Liddle2007MNRAS}. Both criteria balance the goodness of fit and the complexity of the model, penalizing models with more free parameters. Lower AIC or BIC values indicate a more statistically favored model.

First of all, we perform spectral fits for the representative plateau of Burst A, by using four different spectral models. The first model is a single BB, which is written as
\begin{equation}
N(E)=\frac{A_\mathrm{obs}}{D_{10}^2} \times \frac{3.29\times10^{-4}E^2}{\exp(E/k_\mathrm{B}T)-1},
\end{equation}
where $k_\mathrm{B}T$ is the observed temperature in units of keV, $A_{\mathrm{obs}}$ is the observed emission area in km$^{2}$ and $D_{10}$ is the distance to the source in units of 10 kpc. The distance of SGR J1935+2154 with $D\simeq 9\ {\rm kpc}$ is adopted \citep{Zhong2020ApJ}. This model involves the fewest assumptions.
Our analysis shows that the BB temperature fluctuates slightly around a constant value. This temperature variation may result from statistical fluctuations caused by insufficient photon counts, or alternatively, it could stem from the dynamical fluctuations of the fireball. In any case, based on the assumption of a light-curve plateau, we further consider ignoring these fluctuations, directly assuming a constant temperature, and using this temperature to perform a joint fitting of time-resolved spectra during the plateau phase (i.e., the temperatures are tied (shared) across different time intervals). The logarithm of the likelihood for the joint fitting is defined as
\begin{equation}
    \ln \mathcal L = \sum_i \ln L_i(\mathcal D_i, \mathcal M_i(\alpha_i, \beta, ...)),
\end{equation}
where $\mathcal D_i$ is the spectra data in $i$th time interval, $\mathcal M_i(\alpha_i, \beta, ...)$ represents the spectral model, $\alpha_i$ is a time-dependent parameter that varies freely across different time intervals, and $\beta$ is a tied parameter for all intervals. The AIC and BIC values presented in Table \ref{appxtab:5burst_spec_resul} indicate that tying the temperatures for the plateau's time intervals can indeed slightly improve the acceptability of the model.

\begin{figure*}[htbp]
\centering
\subfloat[BB]{\includegraphics[width=0.5\textwidth]{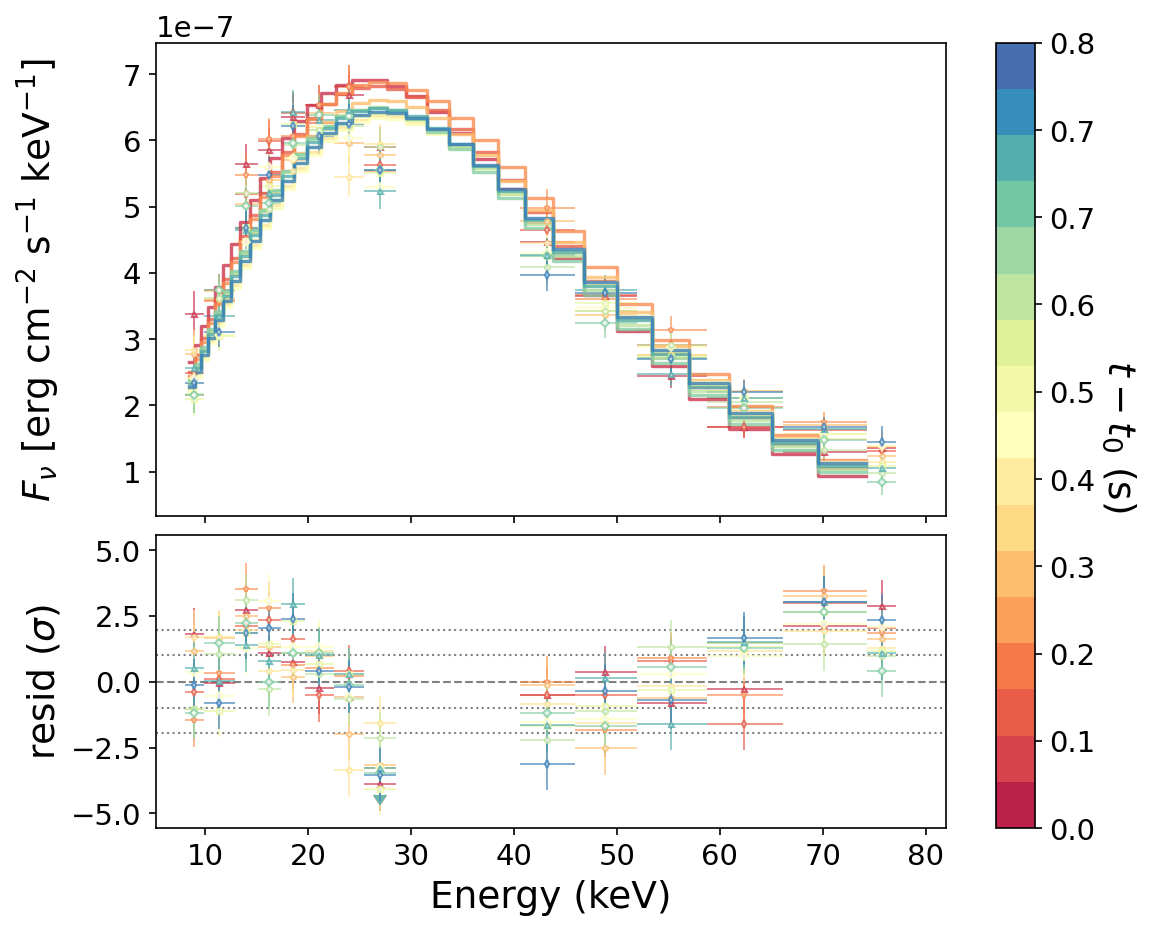}}
\hfill
\subfloat[BB (const. $T$)]{\includegraphics[width=0.5\textwidth]{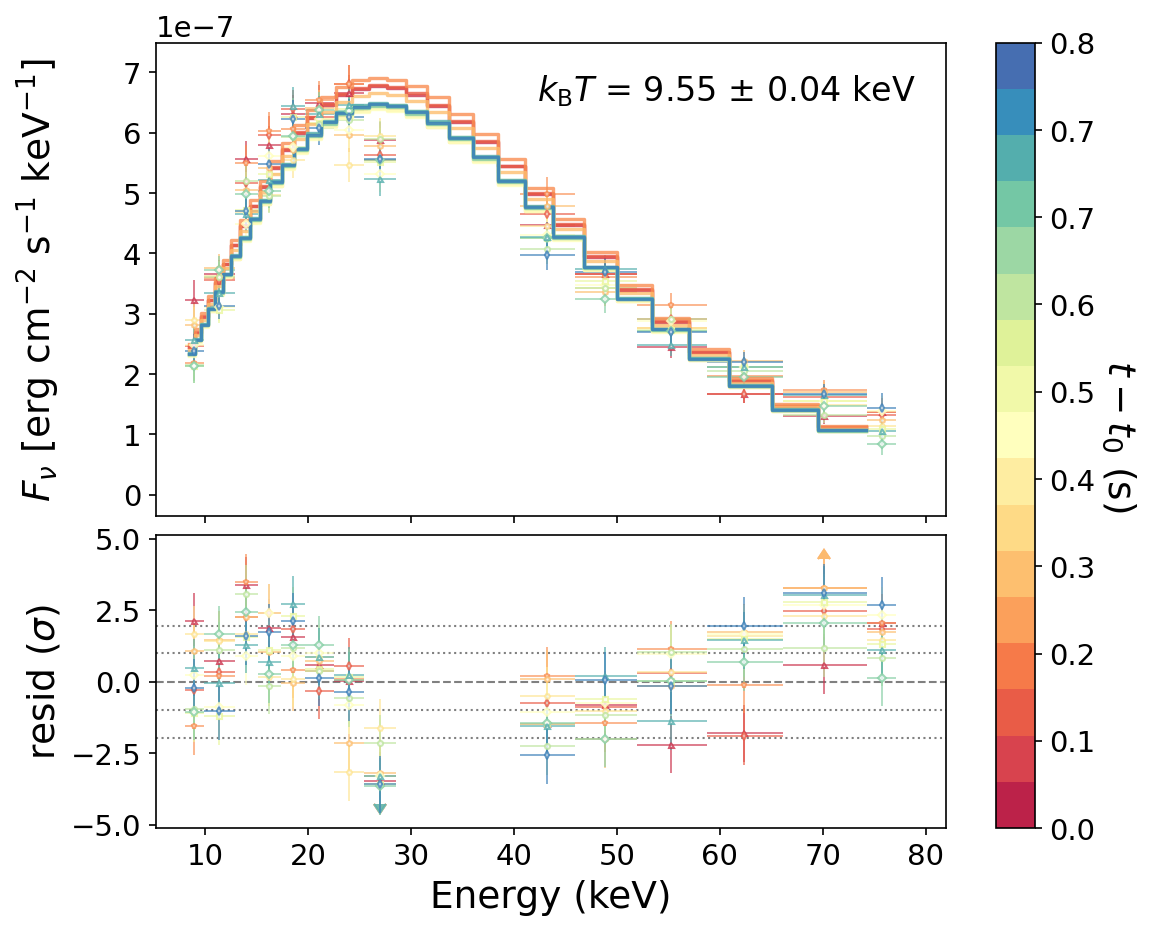}}
\hfill
\subfloat[BB + BB (const. $T$)]{\includegraphics[width=0.5\textwidth]{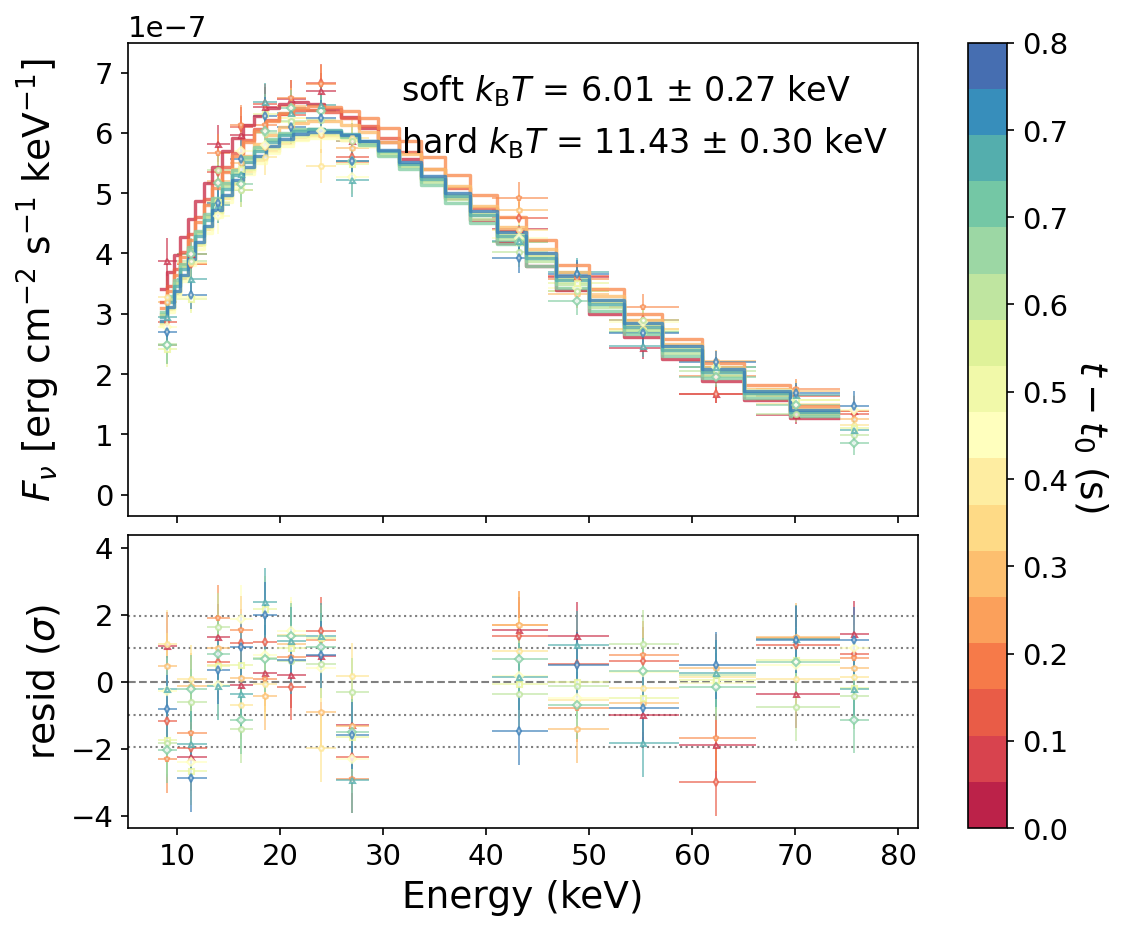}}
\hfill
\subfloat[BB (const. $T$) + CRSF]{\includegraphics[width=0.5\textwidth]{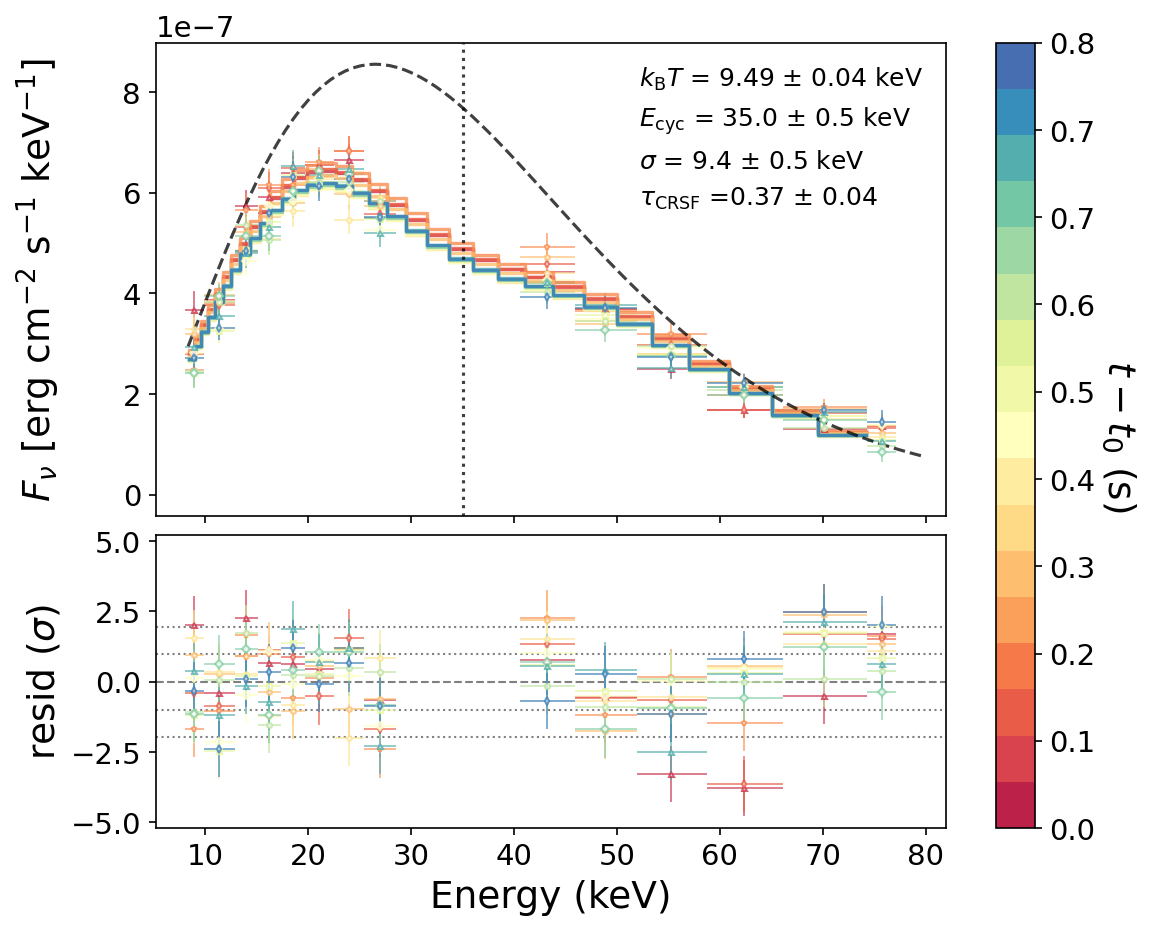}}
\caption{The fits ({\it upper}) and residuals ({\it lower}) of the time-resolved spectra of Burst A during the plateau phase with four different spectral models:
(a) a single BB with free temperature, 
(b) a single BB with a constant temperature for all spectra, 
(c) a double BB with constant temperatures, and 
(d) a single BB with constant temperature plus a CRSF. The dashed line given in panel (d) represents the BB spectrum without the CRSF absorption, where the emission area is taken to be the maximum one. The vertical dotted line represents the center of the CRSF ($E_\mathrm{cyc}$). In the constant-temperature models, the difference between the spectra is solely determined by their different emission areas. }
\label{appxfig:4spec_model_A}
\end{figure*}

However, only with a single BB is there always a significant dip within the energy range from $\sim$20 to $\sim$50 keV, as shown in Figure \ref{appxfig:4spec_model_A}. In order to reduce these discrepancies between the spectral data and model, we further consider another two spectral models by invoking an extra spectral component, which is either another BB or a CRSF absorption. Here the CRSF with a Gaussian profile implemented as a multiplicative model is given by
\begin{equation}
\begin{aligned}
M(E)=\exp\left\{-\tau_{\rm CRSF}\:\exp\left[-\frac{\left(E-E_\mathrm{cyc}\right)^2}{2\sigma^2}\:\right]\right\},
\end{aligned}
\end{equation}
where $E_\mathrm{cyc}$ is the central energy of the line in units of keV, $\sigma$ is the line width, and $\tau_{\rm CRSF}$ is the optical depth at the line center.
Following the same approach as above, the BB temperatures are tied for the time intervals in the plateau phase and the CRSF parameters are tied for all intervals. The results presented in Figure \ref{appxfig:4spec_model_A} show that that both of these scenarios can help to eliminate the dip around $\sim35$ keV. Furthermore, the AIC and BIC values listed in Table \ref{appxtab:5burst_spec_resul} strongly suggest that the best fit can be provided by the BB + CRSF model. The flux data presented in the left panel of Figure \ref{fig:lc_each_plateau} for Burst A are just obtained with this model.

\begin{figure*}[htbp]
\centering
\subfloat[Burst B]{\includegraphics[width=0.5\textwidth]{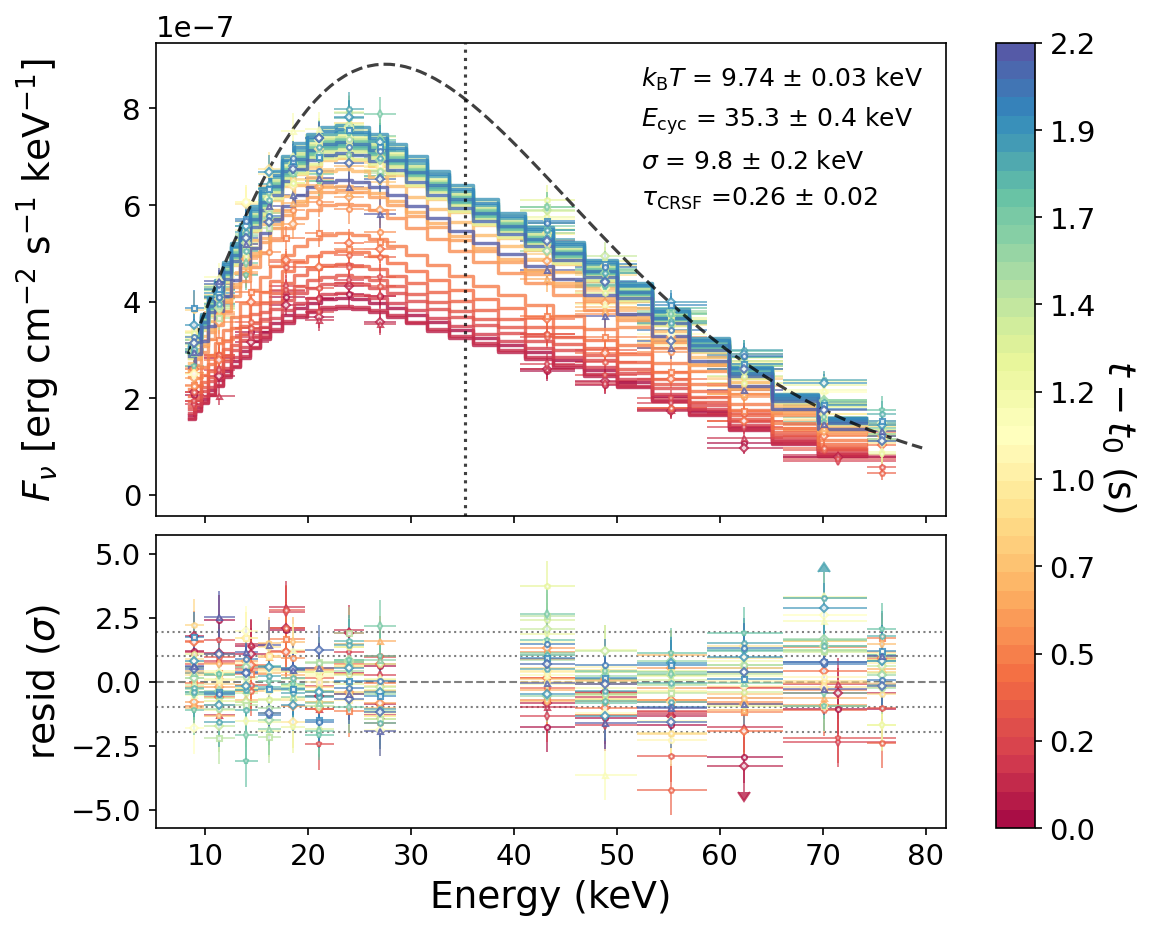}}
\hfill
\subfloat[Burst C]{\includegraphics[width=0.5\textwidth]{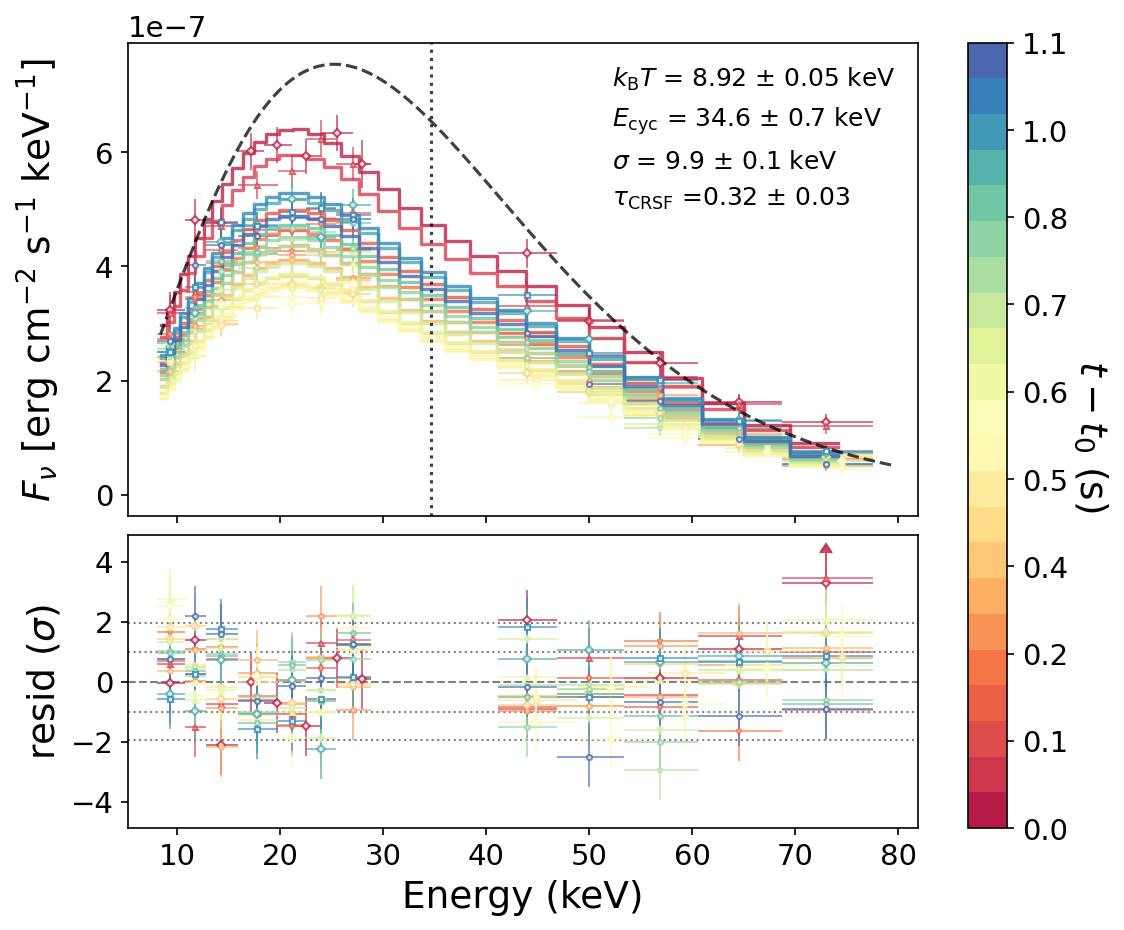}}
\hfill
\subfloat[Burst D]{\includegraphics[width=0.5\textwidth]{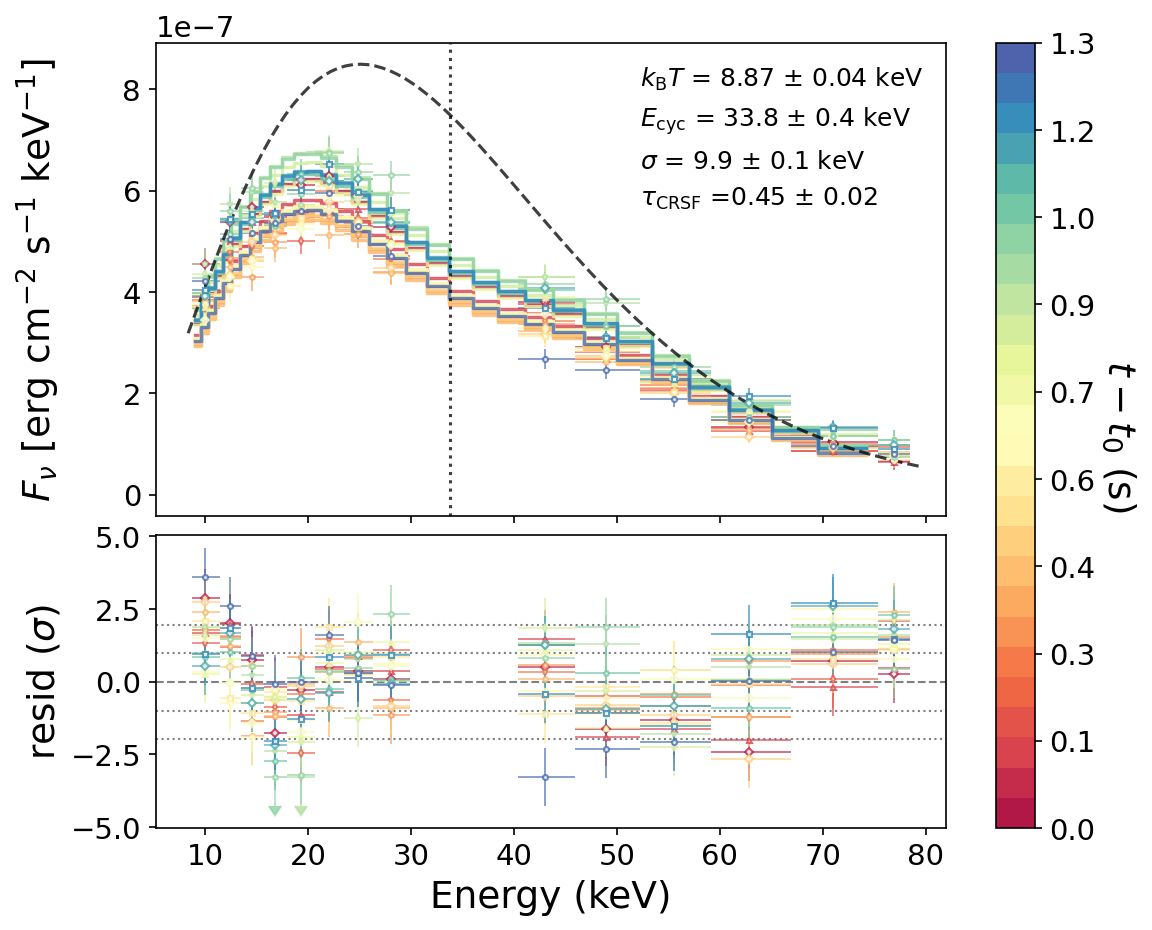}}
\hfill
\subfloat[Burst E]{\includegraphics[width=0.5\textwidth]{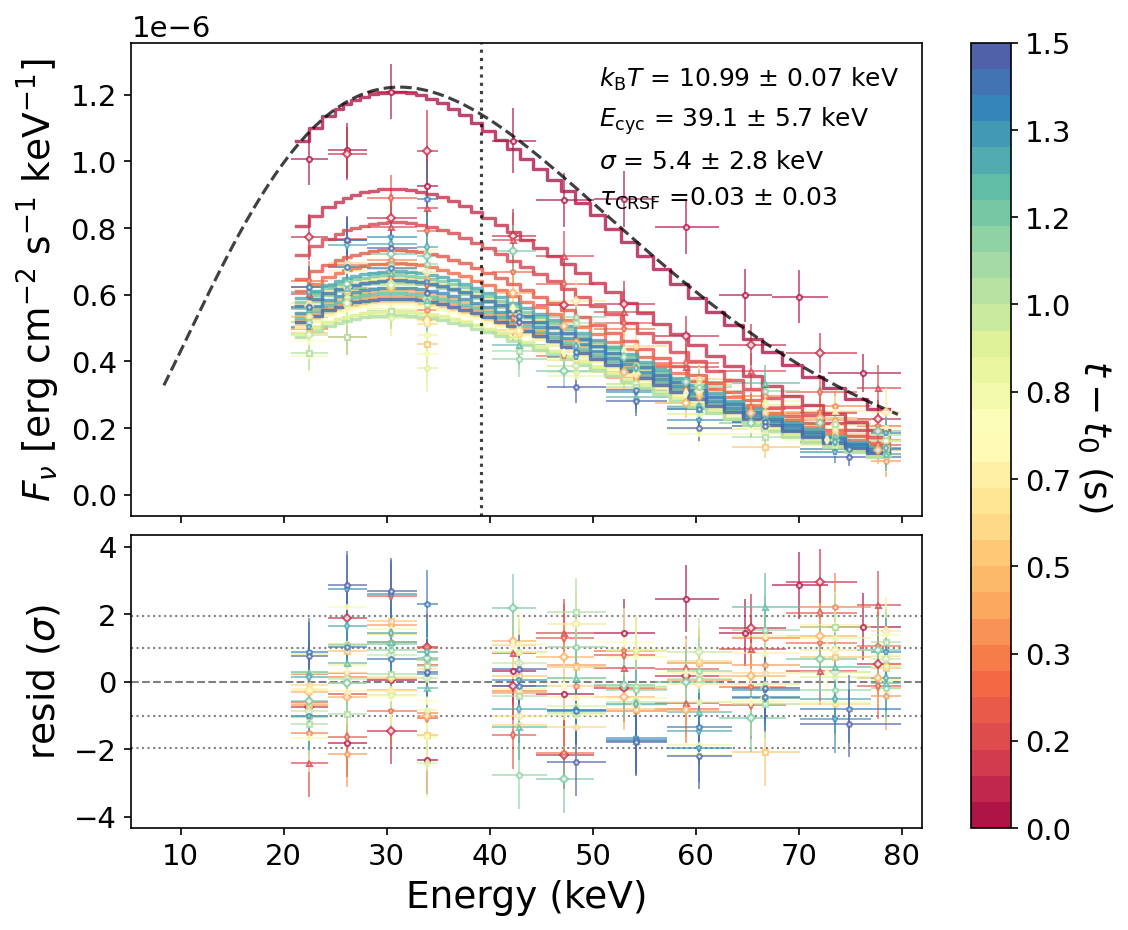}}
\caption{Same as the bottom right panel of Figure \ref{appxfig:4spec_model_A}, but for Bursts B--E. }
\label{appxfig:burst_B2E}
\end{figure*}

Finally, we further apply the BB + CRSF model to fit the time-resolved spectra of all other selected bursts. In fact, we have also examined the other three spectral models, which are, however, disfavored by their AIC and BIC values as listed in Table \ref{appxtab:5burst_spec_resul}. So, for simplicity, we do not present these alternative fitting results. Bursts B--E fittings with the BB + CRSF model are shown in Figure \ref{appxfig:burst_B2E}.
These results demonstrate that the peaks of the time-resolved spectra can always be concentrated around a constant value, and the temporal evolution of the spectra is solely determined by the variation of the emission area.
Figure \ref{fig:lc_each_eclipsed} presents the derived fluxes and emission areas versus time.
The comparison given in the middle panels of this figure further demonstrates that these spectrum-determining emission areas can agree well with the area variation curve derived from the eclipse modeling.


\bibliography{main}{}
\bibliographystyle{aasjournalv7}

\newpage

\newpage
\begin{table}[htbp]
\centering
\caption{The statistical goodness of the spectral fittings of Bursts A--E with four spectral models.}
\label{appxtab:5burst_spec_resul}
\begin{tabular}{llrrr}
\toprule
Burst & Model & $\chi^2$/dof & AIC & BIC \\
\midrule
A & BB & 590.29/196 & 670.49 & 775.18 \\
& BB (const. $T$) & 601.29/206 & 655.14 & 731.80 \\
& BB + BB (const. $T$) & 293.96/182 & 415.95 & 554.99 \\
& BB + CRSF (const. $T$) & 316.99/203 & 378.48 & 463.82 \\
\addlinespace
B & BB & 956.31/495 & 1158.60 & 1503.68 \\
& BB (const. $T$) & 1066.15/533 & 1170.99 & 1371.66 \\
& BB + BB (const. $T$) & 627.41/485 & 857.89 & 1238.42 \\
& BB + CRSF (const. $T$) & 715.28/530 & 827.31 & 1039.89 \\
\addlinespace
C & BB & 477.46/219 & 583.63 & 722.63 \\
& BB (const. $T$) & 499.33/236 & 559.76 & 649.78 \\
& BB + BB (const. $T$) & 253.90/209 & 390.46 & 554.80 \\
& BB + CRSF (const. $T$) & 329.48/233 & 397.50 & 496.64 \\
\addlinespace
D & BB & 1278.93/299 & 1401.42 & 1583.69 \\
& BB (const. $T$) & 1308.46/316 & 1386.46 & 1513.58 \\
& BB + BB (const. $T$) & 359.09/281 & 534.59 & 769.34 \\
& BB + CRSF (const. $T$) & 590.72/313 & 676.22 & 813.43 \\
\addlinespace
E & BB & 345.00/262 & 493.05 & 691.47 \\
& BB (const. $T$) & 396.81/286 & 478.16 & 604.70 \\
& BB + BB (const. $T$) & 392.80/250 & 579.01 & 808.56 \\
& BB + CRSF (const. $T$) & 395.99/283 & 485.06 & 621.20 \\
\bottomrule
\end{tabular}
\end{table}

\end{document}